\documentclass[aip,graphicx]{revtex4}

\pdfoutput=1

\usepackage{color}
\usepackage{graphics}
\usepackage[pdftex]{graphicx}
\usepackage{amsmath}
\usepackage{amsfonts}
\usepackage{float}
\usepackage{natbib}
\usepackage{epstopdf}
\usepackage{verbatim}
\usepackage{algorithm}
\usepackage{algorithmic}
\usepackage{todonotes}
\usepackage{lineno}
\usepackage{bbold}
\usepackage{array}
\usepackage{mathrsfs}
\usepackage{amsbsy}
\usepackage{mathtools}
\usepackage{chngcntr}
\newcolumntype{?}{!{\vrule width 1pt}}
\newcolumntype{C}{>{\centering\arraybackslash}p{3.7cm}}
\newcolumntype{L}{>{\raggedright\arraybackslash}p{3.9cm}}
\newcolumntype{e}{>{\raggedright\arraybackslash}p{2.5cm}}
\newcolumntype{k}{>{\raggedright\arraybackslash}p{2cm}}
\newcolumntype{l}{>{\raggedright\arraybackslash}p{3.2cm}}
\newcolumntype{R}{>{\raggedleft\arraybackslash}p{3.7cm}}

\begin{document}

\title{Computational mechanics of soft filaments}
\author{Mattia Gazzola}
\affiliation{Department of Mechanical Science and Engineering, National Center for Supercomputing Applications, University of Illinois at Urbana-Champaign, Urbana, IL 61801, USA}

\author{Levi H. Dudte}
\affiliation{John A. Paulson School of Engineering and Applied Sciences, Harvard University, Cambridge, MA 02138, USA}

\author{Andrew G. McCormick}
\affiliation{Google, Mountain View, CA 94043, USA}
\affiliation{John A. Paulson School of Engineering and Applied Sciences, Harvard University, Cambridge, MA 02138, USA}

\author{L. Mahadevan}
\email{lm@seas.harvard.edu}
\affiliation{John A. Paulson School of Engineering and Applied Sciences, and Department of Physics, Harvard University, Cambridge, MA 02138, USA}

\begin{abstract}
Soft slender structures are ubiquitous in natural and artificial systems and can be observed at scales that range from the nanometric to the kilometric, from polymers to space tethers. We present a {practical} numerical approach to simulate the dynamics of filaments that, at every cross-section, can undergo all six possible modes of deformation, allowing the filament to {bend, twist, stretch and shear}, {while interacting with complex environments via muscular activity, surface contact, friction and hydrodynamics.} We examine the accuracy of our method by means of several benchmark problems with known analytic solutions. We then demonstrate the capabilities {and robustness} of our approach to solve forward problems in physics and mechanics related to solenoid and plectoneme formation in twisted, stretched filaments, and inverse problems related to active biophysics of limbless locomotion on solid surfaces and in bulk liquids. All together, our approach {provides a robust computational framework to characterize the mechanical response and design of soft active slender structures.}
\end{abstract}


\maketitle

\section{Introduction}
Quasi one-dimensional objects are characterized by having one dimension, the length $L$, much larger than the others, say the radius $r$, so that $L/r \gg 1$. Relative to three-dimensional objects, this measure of slenderness allows for significant mathematical simplification in accurately describing the physical dynamics of strings, filaments and rods. It is thus perhaps not surprising that the physics of strings {has} been the subject of intense study for centuries \cite{Kirchhoff:1859,Clebsch:1883,Love:1906,Cosserat:1909,Antman:1973,Dill:1992,Langer:1996,vanDerHeijden:2000,Goss:2005,Maugin:2010}, and indeed their investigation substantially pre-dates the birth of three-dimensional elasticity.

Following the pioneering work of Galileo on the bending of cantilevers, one-dimensional analytical models of  beams date back to 1761 when Jakob Bernoulli first introduced the use of differential equations to capture the relation between geometry and bending resistance in a \textit{planar elastica}, that is an elastic curve deforming in a two-dimensional space. This attempt was then progressively refined by Huygens, Leibniz and Johann Bernoulli \cite{Goss:2009}, until Euler presented a full solution of the planar elastica, obtained by minimizing the strain energy and by recognizing the relation between flexural stiffness and material and geometric properties. Euler also showed the existence of bifurcating solutions in a rod subject to compression, identifying its first buckling mode, while Lagrange formalized the corresponding multi-modal solution \cite{Antman:1973}. Non-planar deformations of the elastica were first tackled by Kirchhoff \cite{Kirchhoff:1859,Dill:1992} and Clebsch \cite{Clebsch:1883} who envisioned a rod as an assembly of short undeformable straight segments with dynamics determined by contact forces and moments, leading to three-dimensional configurations. Later, Love \cite{Love:1906} approached the problem from a Lagrangian perspective characterizing a filament by contiguous cross sections that can rotate relative to each other, but remain undeformed and perpendicular to the centerline of the rod at all times; in modern parlance this assumption is associated with dynamics on the rotation group SO(3) at every cross-section. The corresponding equations of motion capture the ability of the filament to bend and twist, but not shear or stretch. Eventually the Cosserat brothers \cite{Cosserat:1909} relaxed the assumption of inextensibility and cross-section orthogonality to the centerline, deriving a general mathematical framework that accommodates all six possible degrees of freedom associated with {bending, twisting, stretching and shearing}, effectively formulating dynamics on the full Euclidean group SE(3).

The availability of these strong mathematical foundations \cite{Antman:1973} prompted a number of discrete computational models \cite{Yang:1993,Spillmann:2007,Bergou:2008,Arne:2010,Audoly:2013} that allow for the exploration of a range of physical phenomena. These include, for example, the study of polymers and DNA \cite{Yang:1993,Wolgemuth:2000}, elastic and viscous threads \cite{Bergou:2010,Arne:2010,Audoly:2013,Brun:2015}, botanical applications \cite{Goriely:1998,Gerbode:2012}, elastic ribbons \cite{Bergou:2008}, woven cloth \cite{Kaldor:2008} and tangled hair \cite{Ward:2007}. Because the scaled ratio of the stretching and shearing stiffness to the bending stiffness for slender filaments is $L^2/r^2 \gg 1$, the assumption of inextensibility and unshearability is usually appropriate, justifying the widespread use of the Kirchhoff model in the aforementioned applications. 

However, new technologies such as soft robotics and artificial muscles \cite{Kim:2013,Haines:2014,Park:2016} are generating applications for deformable {and stretchable} elastomeric filamentous structures \cite{Haines:2014} for which the assumption of inextensibility and unshearability is no longer valid. Motivated by these advancements, we move away from the Kirchhoff model in favor of the complete Cosserat theory. We present a {robust and relatively simple} numerical scheme that tracks both the rod centerline and local frame allowing for {bend, twist, stretch and shear} \cite{Cosserat:1909} consistent with the full Euclidean group SE(3), while retaining the Hamiltonian structure of the system and fast discrete operators. This allows us to substantially increase the spectrum of problems amenable to be treated via this class of rod models.

Moving beyond the passive mechanics of {individual} filaments, we also account for the interaction between filaments and complex environments with a number of additional biological and physical features, including muscular activity, self-contact and contact with solid boundaries, isotropic and anisotropic surface friction and viscous interaction with a fluid. Finally, we demonstrate {the  capabilities and the robustness of our solver by embedding it in an inverse design cycle for the identification of optimal terrestrial and aquatic limbless locomotion strategies.}

The paper is structured as follows. In Section \ref{sec:govequations} we review and introduce the mathematical foundations of the model. In Section \ref{sec:discretization} we present the corresponding discrete scheme and validate our framework against a battery of benchmark problems. In Section \ref{sec:additionalphysics} we detail the physical and biological enhancements to the original model, and finally in Section \ref{sec:applications} we showcase the potential of our solver via the study of solenoids and plectonemes as well as limbless biolocomotion. Mathematical derivations and additional validation test cases are presented in the Appendix.

\section{Governing Equations}

\label{sec:govequations}
We consider filaments as slender cylindrical structures deforming in three-dimensions with a characteristic length $L$ which is assumed to be much larger than the radius ($L\gg r$) at any cross section. Then the filament can be geometrically reduced to a one-dimensional representation, and its dynamical behavior may be approximated by averaging all balance laws at every cross section \cite{Antman:1973}. We start with a description of the commonly used Kirchhoff-Love theory that accounts for bend and twist at every cross-section but ignores stretch and shear, before moving on to the Cosserat theory that accounts for these additional degrees of freedom as well.


\subsection{Kirchhoff-Love theory for inextensible, unshearable rods}

\begin{figure}
\begin{center}
\includegraphics[width=\textwidth]{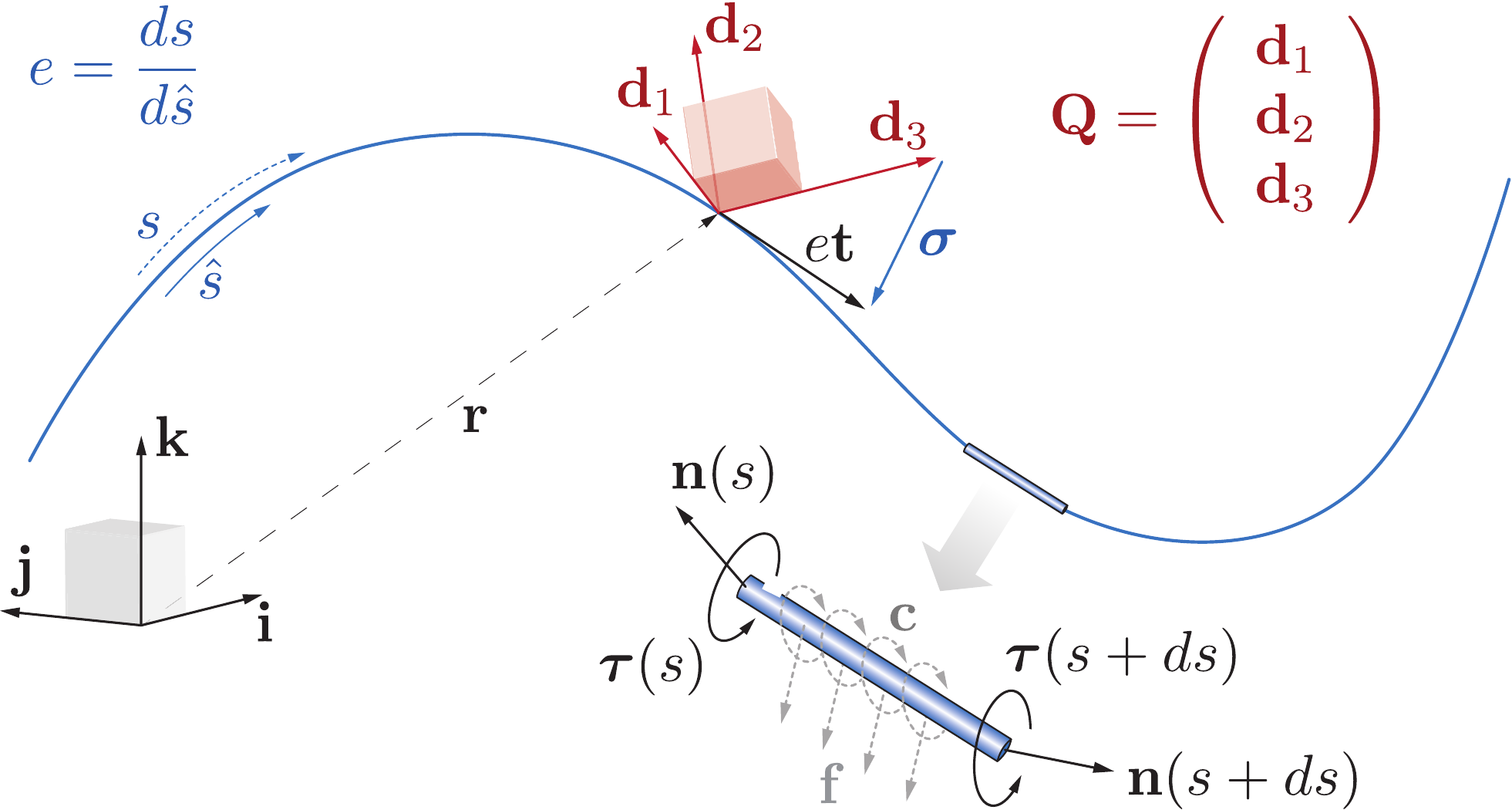}
\caption{\scriptsize{\textbf{The Cosserat rod model.} A filament deforming in the three-dimensional space is represented by a centerline coordinate $\mathbf{r}$ and a material frame characterized by the orthonormal triad $\{\mathbf{d}_1, \mathbf{d}_2, \mathbf{d}_3\}$. The corresponding orthogonal rotation matrix $\mathbf{Q}$ with row entries $\mathbf{d}_1$, $\mathbf{d}_2$, $\mathbf{d}_3$ transforms a vector $\mathbf{x}$ from the laboratory canonical basis $\{\mathbf{i}, \mathbf{j}, \mathbf{k}\}$ to the material frame of reference $\{\mathbf{d}_1, \mathbf{d}_2, \mathbf{d}_3\}$ so that $\mathbf{x}_{\mathcal{L}}=\mathbf{Q}\mathbf{x}$ and vice versa $\mathbf{x}=\mathbf{Q}^T\mathbf{x}_{\mathcal{L}}$. If extension or compression is allowed, the current filament configuration arc-length $s$ may no longer coincide with the rest reference arc-length $\hat{s}$. This is captured via the scalar dilatation field $e=ds/d\hat{s}$. Moreover, to account for shear we allow the triad $\{\mathbf{d}_1, \mathbf{d}_2, \mathbf{d}_3\}$ to detach from the unit tangent vector $\mathbf{t}$ so that $\mathbf{d}_3\neq \mathbf{t}$ (we recall that the condition $\mathbf{d}_3 = \mathbf{t}$ and $e=1$ correspond to the Kirchhoff constraint for unshearable and inextensible rods, and implies that $\boldsymbol{\sigma}=e\mathbf{t}-\mathbf{d}_3=\mathbf{0}$). The dynamics of centerline and material frame are related through quadratic energy functionals that give rise at each cross section to the internal force and torque resultants, $\mathbf{n}$ and $\boldsymbol{\tau}$}. External loads are represented via the {force} $\mathbf{f}$ and {couple} $\mathbf{c}$ line densities.}
\label{fig:cosserat}
\end{center}
\end{figure}

As illustrated in Fig.~\ref{fig:cosserat}, a filament in the Cosserat rod theory can be described by a centerline $\mathbf{r}: (s\in[0,L]\in \mathbb{R},t\in \mathbb{R}^+)\rightarrow\mathbb{R}^3$ and an oriented frame of reference $\mathbf{Q}:(s\in[0,L]\in \mathbb{R},t\in \mathbb{R}^+)\rightarrow\text{SO}(3)$ equivalent to the orthonormal triad of unit vectors $\mathbf{Q} = \{\mathbf{d}_1, \mathbf{d}_2, \mathbf{d}_3\}$. Here, $s$ is the centerline arc-length coordinate in its current configuration and $t$ is time.

Denoting by $\mathbf{x}$ any generic vector represented in the Eulerian frame and $\mathbf{x}_{\mathcal{L}}$ as the body-convected (Lagrangian) frame of reference allows us to write
\begin{eqnarray}
\text{laboratory:}&&~~~~\mathbf{x}=\bar{x}_1\mathbf{i}+\bar{x}_2\mathbf{j}+\bar{x}_3\mathbf{k},\label{eq:eulerianRep}\\
\text{body-convected:}&&~~~~\mathbf{x}_{\mathcal{L}}=x_1\mathbf{d}_1+x_2\mathbf{d}_2+x_3\mathbf{d}_3,\label{eq:lagrangianRep}
\end{eqnarray}
where Eq.~(\ref{eq:eulerianRep}) expresses $\mathbf{x}$ in the laboratory canonical basis $\{\mathbf{i}, \mathbf{j}, \mathbf{k}\}$, while Eq.~(\ref{eq:lagrangianRep}) expresses the same vector in the body-convected director basis $\{\mathbf{d}_1, \mathbf{d}_2, \mathbf{d}_3\}$. Then, the matrix $\mathbf{Q}$ transforms any vector $\mathbf{x}$ from the laboratory to the body-convected representation via $\mathbf{x}_{\mathcal{L}}=\mathbf{Q}\mathbf{x}$ and conversely, $\mathbf{x}=\mathbf{Q}^{-1}\mathbf{x}_{\mathcal{L}}=\mathbf{Q}^{T}\mathbf{x}_{\mathcal{L}}$, since $\mathbf{Q}^T\mathbf{Q}=\mathbf{Q}\mathbf{Q}^T=\pmb{1}$. {In general we need to distinguish between the arc-length coordinate $s$ that corresponds to the current filament configuration {and} the arc-length coordinate $\hat{s}$ associated with the reference configuration of the filament, due to stretching (Fig.~\ref{fig:cosserat} (throughout  we will use hatted quantities to denote the reference configuration).}

 { We will first start by presenting the equations of motion under the assumption of inextensibility (i.e. $s=\hat{s}$), before generalizing them to the stretchable case in the {subsequent} sections. Denoting the rod angular velocity as $\boldsymbol{\omega} = {\rm vec} [ \frac{\partial \mathbf{Q}}{\partial t}^T \mathbf{Q}] $ and the generalized curvature as $\boldsymbol{\kappa} = {\rm vec} [ \frac{\partial \mathbf{Q}}{\partial s}^T \mathbf{Q}]  $, where {$\rm vec [ \mathbf{A} ]$} denotes the 3-vector associated with the skew-symmetric matrix $\mathbf A$,  the following transport identities hold}
\begin{equation}
\mathbf{Q}\frac{\partial \mathbf{x}}{\partial t} = \frac{\partial \mathbf{x}_{\mathcal{L}}}{\partial t} + \boldsymbol{\omega}_{\mathcal{L}}\times \mathbf{x}_{\mathcal{L}},~~~~~~~~~~~~~~
\mathbf{Q}\frac{\partial \mathbf{x}}{\partial s} = \frac{\partial \mathbf{x}_{\mathcal{L}}}{\partial s} + \boldsymbol{\kappa}_{\mathcal{L}}\times \mathbf{x}_{\mathcal{L}}.
\end{equation}
{Using the above equations {(full derivation in the Appendix)} we can express the advection of the rod positions and local frames, as well as the linear and angular momentum balance in a convenient Eulerian-Lagrangian form}
\begin{eqnarray}
\frac{\partial \mathbf{r}}{\partial t} &=& \mathbf{v}\label{eq:vel2} \\
\frac{\partial \mathbf{d}_j}{\partial t} &=& (\mathbf{Q}^T\boldsymbol{\omega}_{\mathcal{L}}) \times \mathbf{d}_j,~~~~~j=1,2,3\label{eq:frame2}\\
\frac{\partial (\rho A \mathbf{v})}{\partial t} &=& \frac{\partial (\mathbf{Q}^T\mathbf{n}_{\mathcal{L}})}{\partial s}+ \mathbf{f}\label{eq:linmoment2}\\
\frac{\partial (\rho\mathbf{I}\boldsymbol{\omega}_{\mathcal{L}})}{\partial t} &=& \frac{\partial \boldsymbol{\tau}_{\mathcal{L}}}{\partial s} + \boldsymbol{\kappa}_{\mathcal{L}}\times\boldsymbol{\tau}_{\mathcal{L}}+ \mathbf{Q}\frac{\partial \mathbf{r}}{\partial s}\times \mathbf{n}_{\mathcal{L}}+(\rho\mathbf{I}\boldsymbol{\omega}_{\mathcal{L}})\times\boldsymbol{\omega}_{\mathcal{L}} + \mathbf{c}_{\mathcal{L}}\label{eq:angmoment2},
\end{eqnarray}
{where $\rho$ is the constant material density, $A$ is the cross sectional area, $\mathbf{v}$ is the velocity, $\mathbf{n}_{\mathcal{L}}$ and $\boldsymbol{\tau}_{\mathcal{L}}$ are respectively the internal force and couple resultants, $\mathbf{f}$ and $\mathbf{c}$ are external body force and torque line densities, and the tensor $\mathbf{I}$ is the second area moment of inertia (throughout this study we assume circular cross sections, see Appendix).} 

To close the above system of equations, Eqs.~(\ref{eq:vel2}-\ref{eq:angmoment2}) and determine the dynamics of the rod, it is necessary to specify the form of the internal forces and torques generated in response to {bend} and twist, corresponding to the three degrees of freedom at every cross-section. The strains are defined as the relative local deformations of the rod with respect to its natural strain-free reference configuration.  Bending and twisting strains are associated with the spatial derivatives of the material frame director field $\{\mathbf{d}_1, \mathbf{d}_2, \mathbf{d}_3\}$ and are characterized by the generalized curvature. Specifically, the components of the curvature projected along the directors ($\boldsymbol{\kappa}_{\mathcal{L}}=\kappa_1\mathbf{d}_1+\kappa_2\mathbf{d}_2+\kappa_3\mathbf{d}_3$) coincides with bending ($\kappa_1, \kappa_2$) and twist ($\kappa_3$) strains in the material frame (Table~\ref{tab:strains}). 

Assuming a linear material constitutive law implies linear \textit{stress-strain} relations. Integration of the torque densities over the cross sectional area $A$ yields the bending and twist rigidities (Table~\ref{tab:strains}), so that the resultant \textit{torque-curvature} relations can be generically expressed in vectorial notation as
\begin{equation}
\boldsymbol{\tau}_{\mathcal{L}}=\mathbf{B}\left(\boldsymbol{\kappa}_{\mathcal{L}}-\boldsymbol{\kappa}^o_{\mathcal{L}}\right), ~~~~~~~~~~
\label{eq:Kconstitutivelaws}
\end{equation}
where $\mathbf{B}\in\mathbb{R}^{3\times3}=\text{diag}$($B_1$, $B_2$, $B_3$) is the {bend}/twist stiffness matrix with $B_1$ the flexural rigidity about $\mathbf{d}_1$, $B_2$ the flexural rigidity about $\mathbf{d}_2$, $B_3$ the twist rigidity about $\mathbf{d}_3$. Here, the vector $\boldsymbol{\kappa}^o_{\mathcal{L}}$ characterizes the intrinsic {curvatures of} a filament that in its stress-free state is not straight. We wish to emphasize here that the constitutive laws are most simply expressed in a local Lagrangian form;  hence the use of $\boldsymbol{\kappa}_{\mathcal{L}}$ and not $\boldsymbol{\kappa}$. 

{The Kirchhoff rod is defined by the additional assumptions that there is  no axial extension or compression or shear strain. Then the arc-length $s$ coincides with $\hat{s}$ at all times, and the tangent to the {centerline} is also normal to the cross-section, so that $\mathbf{t}=\mathbf{d}_3$ \cite{Antman:1973}. This implies that {$\mathbf{n}_{\mathcal{L}}$} serves as a Lagrange multiplier, and that the \textit{torque-curvature} relations of Eq.~\ref{eq:Kconstitutivelaws} are linear.} 

This completes the formulation of the equations of motion for the Kirchhoff rod, and when combined with boundary conditions suffices to  have a well-posed initial boundary value problem. For the general stretchable and shearable case, all geometric quantities (A, $\mathbf{I}$, $\boldsymbol{\kappa}_{\mathcal{L}}$, etc.) must be rescaled appropriately, as addressed in the following sections.

\subsection{Cosserat theory of stretchable and shearable filaments}

{In the general case of soft filaments}, at every cross-section we also wish to capture transverse shear and axial strains in addition to bending and twisting.  Since we wish to account for all six deformation modes associated with the six degrees of freedom at each cross section along the rod, we must augment the Kirchhoff description in the previous section and add three more constitutive laws to define the local stress resultants $\mathbf{n}_{\mathcal{L}}$. In fact, they are no longer defined as Lagrange multipliers that enforce the condition that normals to the cross-section coincide with tangents to the centerline, i.e. we now must have $\mathbf{t} \ne \mathbf{d}_3$. 

The shear and axial strains are associated with the deviations between the unit vector perpendicular to the cross-section and the tangent to the {centerline}, and thus may be {expressed} in terms of the derivatives of the centerline coordinate $\mathbf{r}$.  In the material frame of reference, we characterize these strains by the vector $\boldsymbol{\sigma}$ (Fig.~\ref{fig:cosserat}) which then takes the form
\begin{equation}
\boldsymbol{\sigma}_{\mathcal{L}} = \mathbf{Q}\left(\frac{\partial \mathbf{r}}{\partial \hat{s}}-\mathbf{d}_3\right)=\mathbf{Q}(e\mathbf{t}-\mathbf{d}_3).
\label{eq:sigma}
\end{equation}
Here, the scalar field $e(\hat{s},t)=ds/d\hat{s}$ expresses the local stretching or compression ratio (Fig.~\ref{fig:cosserat}) relative to the rest reference configuration ($\hat{s}$) and $\mathbf{t}$ is the unit tangent vector. 

Whenever the filament undergoes axial stretching or compression the corresponding infinitesimal elements deform and all related geometric quantities are affected. By assuming that the material is incompressible and that the cross sections retain their circular shapes at all times, we can conveniently express the governing equations with respect to the rest reference configuration of the filament (denoted by {a} hat) in terms of the local dilatation $e(\hat{s},t)$. Then, the following relations hold
\begin{equation}
ds=e\cdot d\hat{s},~~~~~A=\frac{\hat{A}}{e},~~~~~\mathbf{I}=\frac{\hat{\mathbf{I}}}{e^2},~~~~~\mathbf{B}=\frac{\hat{\mathbf{B}}}{e^2},~~~~~\mathbf{S}=\frac{\hat{\mathbf{S}}}{e},~~~~~\boldsymbol{\kappa}_{\mathcal{L}}=\frac{\hat{\boldsymbol{\kappa}}_{\mathcal{L}}}{e}.
\label{eq:scaledquantities}
\end{equation}

\begin{table}[!h]
\begin{center}
\begin{tabular}{llll}
\hline
\bf{deformation modes} & ~~\bf{strains} & ~~\bf{rigidities} & ~~\bf{loads} \\
\hline
bending about $\mathbf{d}_1$& ~~$\kappa_1$ & ~~$B_1=EI_1$ & ~~$\tau_1=B_1(\kappa_1-\kappa^o_1)$\\
bending about $\mathbf{d}_2$& ~~$\kappa_2$ & ~~$B_2=EI_2$ & ~~$\tau_2=B_2(\kappa_2-\kappa^o_2)$\\
twist about $\mathbf{d}_3$& ~~$\kappa_3$ & ~~$B_3=GI_3$ & ~~$\tau_3=B_3(\kappa_3-\kappa^o_3)$\\
shear along $\mathbf{d}_1$ & ~~$\sigma_1$ & ~~$S_1=\alpha_c G A$ & ~~$n_1=S_1(\sigma_1-\sigma^o_1)$\\
shear along $\mathbf{d}_2$ & ~~$\sigma_2$ & ~~$S_2=\alpha_c G A$ & ~~$n_2=S_2(\sigma_2-\sigma^o_2)$\\
stretch along $\mathbf{d}_3$ & ~~$\sigma_3$ & ~~$S_3=E A$ & ~~$n_3=S_3(\sigma_3-\sigma^o_3)$\\
\hline
\end{tabular}
\caption{\label{tab:strains} \scriptsize{\textbf{Constitutive laws.} The generalized curvature $\boldsymbol{\kappa}_{\mathcal{L}}$ is associated with the bending $\kappa_1$, $\kappa_2$ about the principal directions ($\mathbf{d}_1$, $\mathbf{d}_2$) and the twist $\kappa_3$ about the longitudinal one ($\mathbf{d}_3$), while $\boldsymbol{\sigma}_{\mathcal{L}} = \mathbf{Q}(e\mathbf{t}-\mathbf{d}_3)$ is associated with the shears $\sigma_1$, $\sigma_2$ along the principal directions ($\mathbf{d}_1$, $\mathbf{d}_2$) and the axial extensional or compression $\sigma_3$ along the longitudinal one ($\mathbf{d}_3$). The material properties of the rod are captured through the Young's ($E$) and shear (G) moduli, while its geometric properties are accounted for via the cross sectional area $A$, the second moment of inertia $\mathbf{I}$ and the constant $\alpha_c=4/3$ for circular cross sections \cite{Gere:2001}. The diagonal entries of the bending/twist $\mathbf{B}\in \mathbb{R}^{3\times3}$ and shear/stretch $\mathbf{S}\in\mathbb{R}^{3\times3}$ matrices are, respectively, ($B_1$, $B_2$, $B_3$) and ($S_1$, $S_2$, $S_3$). Pre-strains are modeled via the intrinsic curvature/twist $\boldsymbol{\kappa}^o_{\mathcal{L}}$ and shear/stretch $\boldsymbol{\sigma}^o_{\mathcal{L}}$.}}
\end{center}
\end{table}

As {with} the Kirchhoff rod, assuming a linear material constitutive law implies linear \textit{stress-strain} relations. Integration of the stress and couple densities over the cross sectional area $A$ yields both the rigidities associated with axial extension and shear (Table~\ref{tab:strains}), so that the resultant \textit{load-strain} relations can be generically expressed in vectorial notation as
\begin{equation}
\mathbf{n}_{\mathcal{L}}=\mathbf{S}\left(\boldsymbol{\sigma}_{\mathcal{L}}-\boldsymbol{\sigma}^o_{\mathcal{L}}\right),
\label{eq:constitutivelaws}
\end{equation}
where   $\mathbf{S}\in\mathbb{R}^{3\times3}=\text{diag}$($S_1$, $S_2$, $S_3$) is the shear/stretch stiffness matrix with $S_1$ the shearing rigidity along $\mathbf{d}_1$, $S_2$ the shearing rigidity along $\mathbf{d}_2$, and $S_3$ the axial rigidity along $\mathbf{d}_3$. Here, as {with} the Kirchhoff rod, the vector $\boldsymbol{\sigma}^o_{\mathcal{L}}$ corresponds to the intrinsic shear and stretch, and must be accounted for in the case of stress-free shapes that are non-trivial. {Although the intrinsic strains $\boldsymbol{\kappa}^o_{\mathcal{L}}$, $\boldsymbol{\sigma}^o_{\mathcal{L}}$ are implemented in our solver to account for pre-strained configurations, to simplify the notation in the remaining text we will assume that the filament is intrinsically straight in a stress-free state, so that} $\boldsymbol{\sigma}^o_{\mathcal{L}}=\boldsymbol{\kappa}^o_{\mathcal{L}}=\mathbf{0}$.

The rigidities associated with {bending, twisting, stretching and shearing} are specified in Table 1, and can be expressed as the product of a material component, represented by the Young's ($E$) and shear ($G$) moduli, and a geometric component represented by $A$, $\mathbf{I}$ and the constant $\alpha_c=4/3$ for circular cross sections \cite{Gere:2001}. We note that the rigidity matrices $\mathbf{B}$ and $\mathbf{S}$ are assumed to be diagonal throughout this study, although off diagonal entries can be easily accommodated to model anisotropic materials such as composite elements. In general, this mathematical formulation can be extended to tackle a richer set of physical problems including viscous threads \cite{Arne:2010, Audoly:2013}, magnetic filaments \cite{Landau:1984}, etc., by simply modifying the entries of $\mathbf{B}$ and $\mathbf{S}$ and introducing time-dependent constitutive laws wherein $\boldsymbol{\tau}_{\mathcal{L}}(\boldsymbol{\kappa}_{\mathcal{L}}, \partial_t\boldsymbol{\kappa}_{\mathcal{L}})$ and $\mathbf{n}_{\mathcal{L}}(\boldsymbol{\sigma}_{\mathcal{L}}, \partial_t \boldsymbol{\sigma}_{\mathcal{L}})$, as for example in  \cite{Arne:2010}. We also emphasize here that in the case of stretchable rods, $A$ and $\mathbf{I}$ are no longer constant, rendering the \textit{load-strain} relations non-linear (Eqs.~\ref{eq:Kconstitutivelaws}, \ref{eq:constitutivelaws}), even though the \textit{stress-strain} relations remain linear.

{Having generalized the constitutive relations to account for filament stretchiness and shearability, we now generalize the equations of motion for this case. Multiplying both sides of Eqs.~(\ref{eq:linmoment2},\ref{eq:angmoment2}) by $ds$ and substituting the above identities together with the constitutive laws of Eqs.~(\ref{eq:constitutivelaws}) into Eqs.~(\ref{eq:vel2}-\ref{eq:angmoment2}), yields the final system}
\begin{eqnarray}
\frac{\partial \mathbf{r}}{\partial t} &=& \mathbf{v}\label{eq:velfinal} \\
\frac{\partial \mathbf{d}_j}{\partial t} &=& (\mathbf{Q}^T\boldsymbol{\omega}_{\mathcal{L}}) \times \mathbf{d}_j,~~~~~j=1,2,3\label{eq:framefinal}\\
dm \cdot \frac{\partial^2 \mathbf{r}}{\partial t ^2} &=& \underbrace{\frac{\partial}{\partial \hat{s}} \left(\frac{\mathbf{Q}^T\hat{\mathbf{S}}\boldsymbol{\sigma}_{\mathcal{L}}}{e}\right)d\hat{s}}_{\text{{shear/stretch} internal force}} +\underbrace{\mathbf{F}\label{eq:linmomentfinal}}_{\text{ext. force}}\\
\frac{d\hat{\mathbf{J}}}{e} \cdot \frac{\partial \boldsymbol{\omega}_{\mathcal{L}}}{\partial t} &=& \underbrace{\frac{\partial}{\partial \hat{s}}\left(\frac{\hat{\mathbf{B}}\hat{\boldsymbol{\kappa}}_{\mathcal{L}}}{e^3}\right)d\hat{s} + \frac{\hat{\boldsymbol{\kappa}}_{\mathcal{L}}\times\hat{\mathbf{B}}\hat{\boldsymbol{\kappa}}_{\mathcal{L}}}{e^3}d\hat{s}}_{\text{{bend/twist} internal couple}}~~~+ \underbrace{\left(\mathbf{Q}\mathbf{t}\times\hat{\mathbf{S}}\boldsymbol{\sigma}_{\mathcal{L}}\right)d\hat{s}}_{\text{{shear/stretch} internal couple}}\nonumber\\
&&+\underbrace{\left(d\hat{\mathbf{J}}\cdot\frac{\boldsymbol{\omega}_{\mathcal{L}}}{e}\right)\times \boldsymbol{\omega}_{\mathcal{L}}}_{\text{Lagrangian transport}} + \underbrace{\frac{d\hat{\mathbf{J}}\boldsymbol{\omega}_{\mathcal{L}}}{e^2}\cdot\frac{\partial e}{\partial t}}_{\text{unsteady dilatation}} + \underbrace{\mathbf{C}_{\mathcal{L}}}_{\text{ext. couple}},\label{eq:angmomentfinal}
\end{eqnarray}
where $dm=\rho \hat{A} d\hat{s}=\rho A ds$ is the infinitesimal mass element, and $d\hat{\mathbf{J}}=\rho\hat{\mathbf{I}} d\hat{s}$ is the infinitesimal mass second moment of inertia. We note that the left hand side and the unsteady dilatation term of Eq.~(\ref{eq:angmomentfinal}) arise from the expansion of the original rescaled angular momentum $d\hat{\mathbf{J}}\cdot\partial_t(\boldsymbol{\omega}_{\mathcal{L}}/e)$ via chain rule. We also note that the external force and couple are defined as $\mathbf{F}=e\mathbf{f}d\hat{s}$ and $\mathbf{C}_{\mathcal{L}}=e\mathbf{c}_{\mathcal{L}}d\hat{s}$ (with $\mathbf{f}$ and $\mathbf{c}_{\mathcal{L}}$ the force and torque line densities, respectively) so as to account for the {dependence} on $e$.

Combined with some initial and boundary conditions, Eqs.~(\ref{eq:velfinal}-\ref{eq:angmomentfinal}) express the dynamics and kinematics of the Cosserat rod with respect to its initial rest configuration, in a form suitable to be discretized as described in Section~\ref{sec:discretization}.

\section{Numerical Method}
\label{sec:discretization}

To derive the numerical method for the time evolution of a filament in analogy with the continuum model of Section \ref{sec:govequations}, we first recall a few useful definitions for effectively implementing rotations. We then present the {spatially discretized} model of the rod, and the time discretization approach employed to evolve the governing equations.

\begin{figure}
\begin{center}
\includegraphics[width=\textwidth]{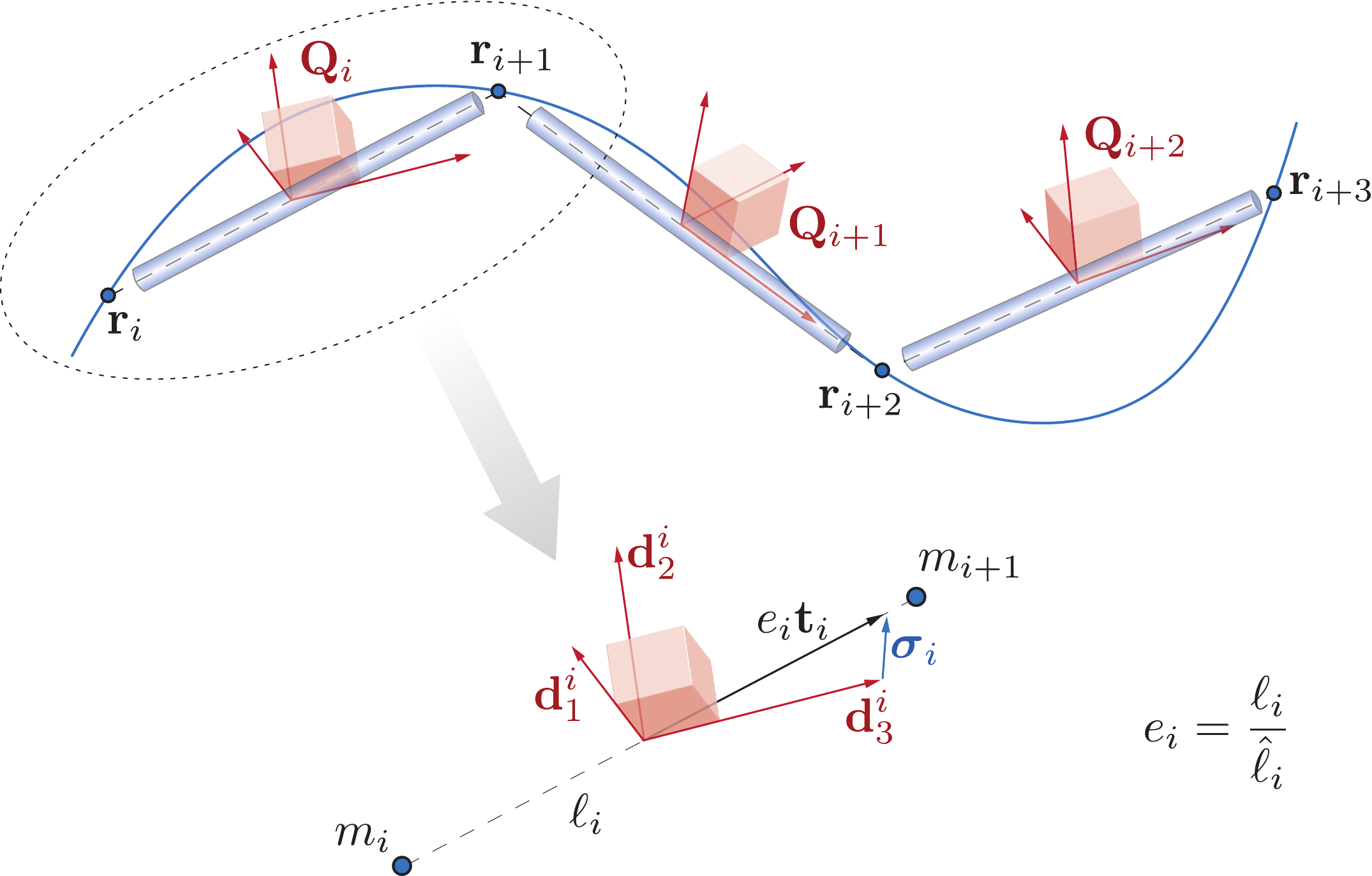}
\caption{\scriptsize{\textbf{Discretization model.} A discrete filament is represented through a set of vertices $\mathbf{r}(t)_{i=1,\dots,N+1}$ and a set of material frames $Q_i(t)=\{\mathbf{d}_1^i,\mathbf{d}_2^i,\mathbf{d}_3^i,\}_{i=1,\dots,N}$. Two consecutive vertices define an edge of length $\ell_i$ along the tangent unit vector $\mathbf{t}_i$. The dilatation is defined as $e_i=\ell_i/\hat{\ell}_i$, where $\hat{\ell}_i$ is the edge rest length. The vector $\boldsymbol{\sigma}_i = e_i\mathbf{t}_i-\mathbf{d}_3^i$ represents the discrete shear and axial strains. The mass $m_r$ of the filament is discretized in pointwise concentrated masses $m_{i=1,\dots,n+1}$ at the locations $\mathbf{r}_i$ for the purpose of advecting the vertices in time. For the evolution of $\mathbf{Q}_i$ in time, we consider instead the mass second moment of inertia $\hat{\mathbf{J}}_{i=1,\dots,n}$ associated with the cylindrical elements depicted in blue.}}
\label{fig:dicretizedmodel}
\end{center}
\end{figure}

\subsection{Rotations}
\label{sec:rotations}
{Bending and twisting deformations of a filament involve} rotations of its material frame $\mathbf{Q}$ in space and time. To numerically simulate the rod, it is critical to represent and efficiently compute these nonlinear geometric transformations fast and accurately. A convenient way to express rotations in space or time is the matrix exponential \cite{Baillieul:1987,Levi:1996,Goriely:2000}. Assuming that the matrix $\mathbf{R}$ denotes the rotation by the angle $\theta$ about the unit vector axis $\mathbf{u}$, then this rotation can be expressed through the exponential matrix $\mathbf{R}=e^{\theta\mathbf{u}}:\mathbb{R}^3\rightarrow\mathbb{R}^{3\times3}$, and efficiently computed via the Rodrigues formula \cite{Rodrigues:1840}
\begin{equation}
e^{\theta\mathbf{u}}=\pmb{1}+\sin \theta\mathbf{U}+(1-\cos \theta)\mathbf{U}\mathbf{U}.
\end{equation}
Here $\mathbf{U}\in\mathbb{R}^{3\times3}$ represents the skew-symmetric matrix associated with the unit vector $\mathbf{u}$
\[
\mathbf{U} = [\mathbf{u}]_{\times}=\left( \begin{array}{ccc}
0 & -u_3 & u_2 \\
u_3 & 0 & -u_1 \\
-u_2 & u_1 & 0 \end{array} \right),
~~~~~~~\mathbf{u} = [\mathbf{U}]^{-1}_{\times}=\left( \begin{array}{c}
~~U_{3,2}\\
-U_{3,1} \\
~~U_{2,1}
\end{array} \right),
\]
where the operator $[\cdot]_{\times}:\mathbb{R}^3\rightarrow\mathbb{R}^{3\times3}$ allows us to transform a vector into the corresponding {skew-symmetric} matrix, and {vice versa} $[\cdot]^{-1}_{\times}:\mathbb{R}^{3\times3}\rightarrow\mathbb{R}^3$.

Conversely, given a rotation matrix $\mathbf{R}$, the corresponding rotation vector can be directly computed via the matrix logarithm operator $\log(\cdot):\mathbb{R}^{3\times3}\rightarrow\mathbb{R}^3$
\[
\theta\mathbf{u}=\log\left(\mathbf{R}\right)= 
\begin{dcases}
    \mathbf{0}& \text{if } \theta=0\\
    \frac{\theta}{2\sin\theta}\left[\mathbf{R}-\mathbf{R}^T\right]^{-1}_{\times}              & \text{if } \theta\neq0, \theta\in(-\pi,\pi)
\end{dcases},
~~~~~~~~\theta=\arccos\left(\frac{tr\mathbf{R}-1}{2}\right).
\]

It is important to notice that the rotation axis $\mathbf{u}$ is expressed in the material frame of reference associated with the matrix $\mathbf{R}$ (or $\mathbf{Q}$). With these tools in hand, we now proceed to outline our numerical scheme.

\subsection{Spatial discretization}
Drawing from previous studies of unshearable and inextensible rods \cite{Spillmann:2007,Bergou:2008,Sobottka:2008}, we capture the deformation of a filament in three-dimensional space via the time evolution of a discrete set of vertices $\mathbf{r}_i(t)\in\mathbb{R}^3, i \in [1,n+1]$ and a discrete set of material frames $\mathbf{Q}_i(t)\in\mathbb{R}^{3\times 3}, i \in [1,n]$, as illustrated in Fig.~\ref{fig:dicretizedmodel}.

Each vertex is associated with the following discrete quantities
\begin{eqnarray}
\mathbf{r}_{i=1,\dots,n+1} &\rightarrow& \mathbf{v}_i=\frac{\partial \mathbf{r}_i}{\partial t},~~~~m_i,~~~~\mathbf{F}_i,
\end{eqnarray}
where $\mathbf{v}_i$ is the velocity , $m_i$ is a pointwise concentrated mass, and $\mathbf{F}_i$ is the external force given in Eq.~(\ref{eq:linmomentfinal}).

Each material frame is associated with an edge $\boldsymbol{\ell}_i$ connecting two consecutive vertices, and with the related discrete quantities
\begin{eqnarray}
\mathbf{Q}_{i=1,\dots,n} &\rightarrow& \boldsymbol{\ell}_i=\mathbf{r}_{i+1}-\mathbf{r}_{i},~~~~\ell_i=|\boldsymbol{\ell}_i|,~~~~\hat{\ell}_i=|\hat{\boldsymbol{\ell}}_i|,~~~~e_i=\frac{\ell_i}{\hat{\ell}_i},~~~~
{\mathbf{t}_i=\frac{\boldsymbol{\ell}_i}{\ell_i}},\nonumber\\
&&\boldsymbol{\sigma}^i_{\mathcal{L}}=\mathbf{Q}_i(e_i\mathbf{t}_i-\mathbf{d}^3_i),~~~~\boldsymbol{\omega}^i_{\mathcal{L}},~~~~\hat{A}_i,~~~~\hat{\mathbf{J}}_i,~~~~\hat{\mathbf{B}}_i,~~~~\hat{\mathbf{S}}_i,~~~~\mathbf{C}^i_{\mathcal{L}},
\end{eqnarray}
where $\ell_i=|\boldsymbol{\ell}_i|$, $\hat{\ell}_i=|\hat{\boldsymbol{\ell}}_i|$, $e_i=\ell_i / \hat{\ell}_i$ are the edge current length, reference length and dilatation factor, $\mathbf{t}_i$ is the discrete tangent vector, $\boldsymbol{\sigma}^i_{\mathcal{L}}$ is the discrete shear/axial strain vector, $\boldsymbol{\omega}^i_{\mathcal{L}}$ is the discrete angular velocity,  $\hat{A}_i$, $\hat{\mathbf{J}}_i$, $\hat{\mathbf{B}}_i$, $\hat{\mathbf{S}}_i$ are the edge reference {cross section} area, mass second moment of inertia, {bend/twist} matrix and shear/stretch matrix, and finally $\mathbf{C}^i_{\mathcal{L}}$ is the external couple given in Eq.~(\ref{eq:angmomentfinal}).

Whereas in the continuum setting (Section \ref{sec:govequations}) all quantities are defined pointwise, in a discrete setting some quantities, and in particular $\boldsymbol{\kappa}_{\mathcal{L}}$, are naturally expressed in an integrated form over the domain $\boldsymbol{\mathcal{D}}$ along the filament \cite{Grinspun:2006,Bergou:2008}. Any integrated quantity divided by the corresponding integration domain length $\mathcal{D}=|\boldsymbol{\mathcal{D}}|$ is equivalent to its pointwise average. In the context of our discretization the domain $\boldsymbol{\mathcal{D}}$ becomes the Voronoi region $\boldsymbol{\mathcal{D}}_i$ of length
\begin{equation}
\mathcal{D}_i=\frac{\ell_{i+1}+\ell_i}{2},
\end{equation}
which is defined only for the \textit{interior} vertices $\mathbf{r}^{(int)}_{i=1,\dots,n-1}$. Each interior vertex is then also associated with the following discrete quantities
\begin{eqnarray}
\mathbf{r}^{(int)}_{i=1,\dots,n-1} &\rightarrow& \mathcal{D}_i,~~~\hat{\mathcal{D}}_i,~~~\mathcal{E}_i=\frac{\mathcal{D}_i}{\hat{\mathcal{D}}_i},~~~\hat{\boldsymbol{\kappa}}^{i}_{\mathcal{L}}=\frac{\log(\mathbf{Q}_{i+1}\mathbf{Q}^T_{i})}{\hat{\mathcal{D}}_i},~~~\hat{\boldsymbol{\mathcal{B}}}_i=\frac{\hat{\mathbf{B}}_{i+1}\hat{\ell}_{i+1}+\hat{\mathbf{B}}_i\hat{\ell}_i}{2\hat{\mathcal{D}}_i},
\end{eqnarray}
where $\hat{\mathcal{D}}_i$ is the Voronoi domain length at rest and $\mathcal{E}_i$ is Voronoi region dilatation factor. Recalling that the generalized curvature expresses a rotation per unit length about its axis, then the quantity $\hat{\mathcal{D}}_i\hat{\boldsymbol{\kappa}}^i_{\mathcal{L}}$ naturally expresses the rotation that transforms a material frame $\mathbf{Q}_{i}$ to its neighbour $\mathbf{Q}_{i+1}$ over the segment size $\hat{\mathcal{D}}_i$ along the rod. Therefore, the relation $e^{\hat{\mathcal{D}}_i\hat{\boldsymbol{\kappa}}^i_{\mathcal{L}}}\mathbf{Q}_{i}=\mathbf{Q}_{i+1}$ holds, so that $\hat{\boldsymbol{\kappa}}^i_{\mathcal{L}}=\log(\mathbf{Q}_{i+1}\mathbf{Q}^T_{i})/\hat{\mathcal{D}}_i$. Finally, we introduce the {bend/twist} stiffness matrix $\hat{\boldsymbol{\mathcal{B}}}_i$ consistent with the Voronoi representation.

Then, we may discretize the governing Eqs.~(\ref{eq:velfinal}-\ref{eq:angmomentfinal}) so that they read
\begin{eqnarray}
\frac{\partial \mathbf{r}_i}{\partial t} &=& \mathbf{v}_i,\hspace{7.5cm}i=[1,n+1]\label{eq:velfinaldiscr}\\
\frac{\partial \mathbf{d}_{i,j}}{\partial t} &=& (\mathbf{Q}_i^T\boldsymbol{\omega}^i_{\mathcal{L}}) \times \mathbf{d}_{i,j},\hspace{4.5cm}i=[1,n], ~j=1,2,3\label{eq:framefinaldiscr}\\
m_i \cdot \frac{\partial \mathbf{v}_i}{\partial t} &=& \Delta^h \left(\frac{\mathbf{Q}_i^T\hat{\mathbf{S}}_i\boldsymbol{\sigma}^i_{\mathcal{L}}}{e_i}\right) +\mathbf{F}_i,\hspace{4.5cm}i=[1,n+1]\label{eq:linmomentfinaldiscr}\\
\frac{\hat{\mathbf{J}}_i}{e_i} \cdot \frac{\partial \boldsymbol{\omega}^i_{\mathcal{L}}}{\partial t} &=& \Delta^h\left(\frac{\hat{\boldsymbol{\mathcal{B}}}_i\hat{\boldsymbol{\kappa}}_{\mathcal{L}}^{i}}{\mathcal{E}_i^3}\right) + \mathcal{A}^h\left(\frac{\hat{\boldsymbol{\kappa}}_{\mathcal{L}}^i\times\hat{\boldsymbol{\mathcal{B}}}_i\hat{\boldsymbol{\kappa}}_{\mathcal{L}}^i}{\mathcal{E}_i^3}\hat{\mathcal{D}}_i\right) + \left(\mathbf{Q}_i\mathbf{t}_i\times\hat{\mathbf{S}}_i\boldsymbol{\sigma}^i_{\mathcal{L}}\right)\hat{\ell}_i\nonumber\\
&&+\left(\hat{\mathbf{J}}_i\cdot\frac{\boldsymbol{\omega}^i_{\mathcal{L}}}{e_i}\right)\times \boldsymbol{\omega}^i_{\mathcal{L}} + \frac{\hat{\mathbf{J}}_i\boldsymbol{\omega}^i_{\mathcal{L}}}{e_i^2}\cdot\frac{\partial e_i}{\partial t} + \mathbf{C}^i_{\mathcal{L}},\hspace{2.5cm}i=[1,n]\label{eq:angmomentfinaldiscr}
\end{eqnarray}
{where $\Delta^h:\{\mathbb{R}^3\}_N\rightarrow\{\mathbb{R}^3\}_{N+1}$ is the discrete difference operator and $\mathcal{A}^h:\{\mathbb{R}^3\}_N\rightarrow\{\mathbb{R}^3\}_{N+1}$ is the averaging operator to transform integrated quantities over the domain $\mathcal{D}$ to their point-wise counterparts. We note that $\Delta^h$ and $\mathcal{A}^h$ {operate} on a set of $N$ vectors and returns $N+1$ vectors, consistent with Eqs.~(\ref{eq:velfinaldiscr}-\ref{eq:angmomentfinaldiscr}) (see the Appendix for further details).}

\subsection{Time discretization}

{While the above equations are conservative and preserve energy, in general when additional physical effects and interactions with complex environments are considered, the equations of motion are not conservative. We choose a symplectic, second-order Verlet scheme to integrate the equations of motion in time so that our numerical scheme is energy-preserving in the case of conservative dynamics. We note that despite the failure of Verlet schemes to integrate rotational equations of motion when represented by quaternions, in our case their use is acceptable as rotations are represented instead by Euler angles \cite{Kol:1997}.}

{The second order position Verlet time integrator is structured in three blocks: a first half-step updates the linear and angular positions, followed by the evaluation of local linear and angular accelerations, and finally a second half-step updates the linear and angular positions again. Therefore, it entails only one right hand side evaluation of Eqs.~(\ref{eq:linmomentfinaldiscr}, \ref{eq:angmomentfinaldiscr}), the most computationally expensive operation (see Appendix for details).}

This algorithm strikes a balance between computing costs, numerical accuracy and implementation modularity: by foregoing an implicit integration scheme we can incorporate a number of additional physical effects and soft constraints, even though this may come at the expense of computational efficiency.

\subsection{Validation}

\label{sec:validation}

We first validate our proposed methodology against a number of benchmark problems with analytic solutions and examine the convergence properties of our approach. {Three case studies serve to characterize the competition between bending and twisting effects in the context of helical buckling, dynamic stretching of a loaded rod under gravity, and the competition between shearing and bending in the context of a Timoshenko beam. Further validations reported in the Appendix include Euler and Mitchell buckling due to compression or twist, and stretching and twisting vibrations.}


\subsubsection{Helical buckling instability}
\label{sec:helical}

\begin{figure}
\begin{center}
\includegraphics[width=0.9\textwidth]{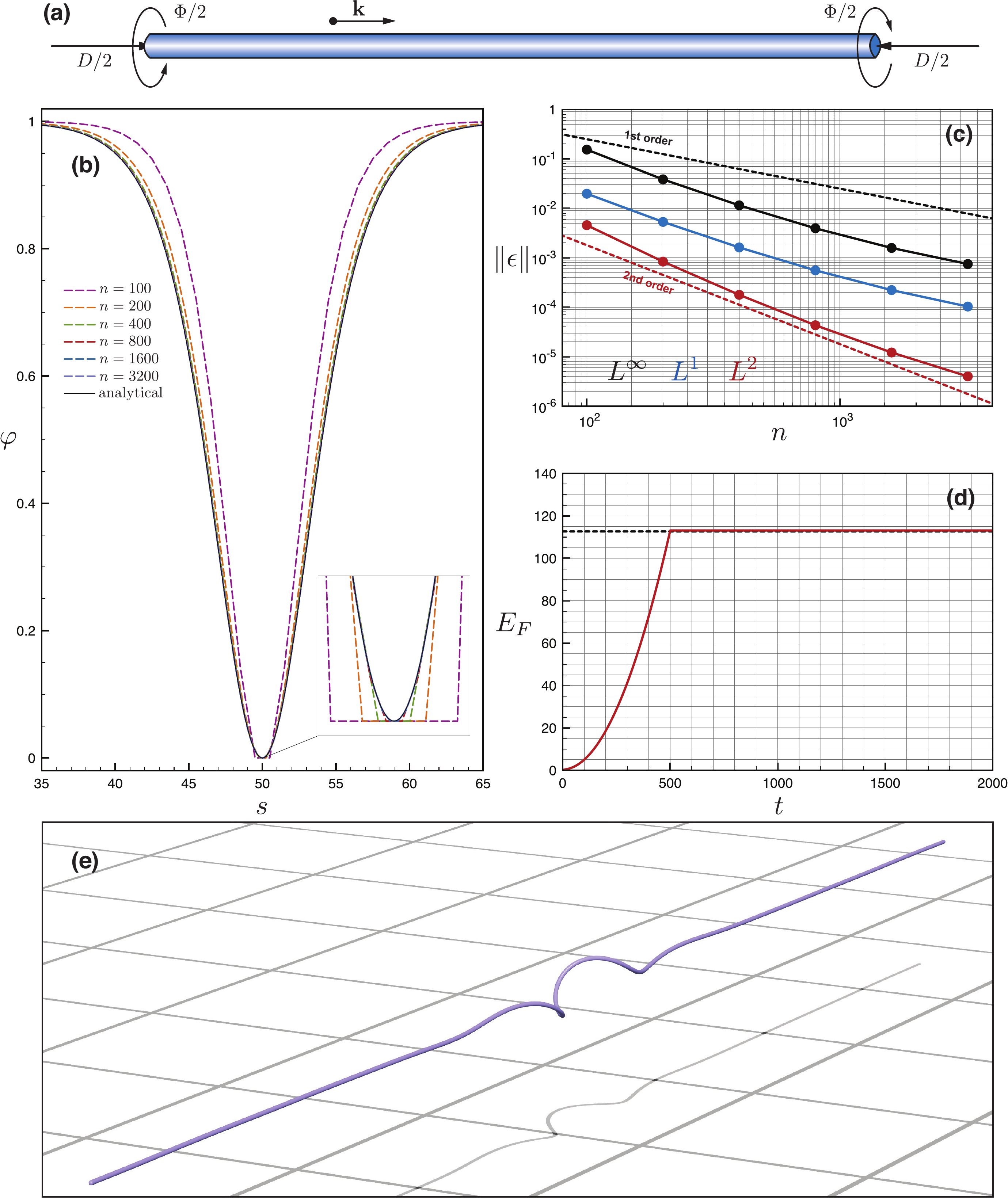}
\caption{\scriptsize{\textbf{Time-space convergence study for localized helical buckling.} (a) We consider a rod originally straight whose ends are pulled together in the axial direction $\mathbf{k}$ with a slack $D/2$ and simultaneously twisted by the angle $\Phi/2$. (b) Comparison between the analytical envelope function $\varphi(s)$ and numerical approximations $\varphi^n(s)$ at different levels of time-space resolution. Here, the time discretization $\delta t$ is slaved by the spatial discretization $\delta l = L/n$ according to $\delta t = 10^{-3} \delta l$ s. (c) Norms $L^{\infty}(\epsilon)$ (black), $L^1(\epsilon)$ (blue) and $L^2(\epsilon)$ (red) are plotted against the number of discretization elements $n$. (d) Time evolution of the total energy of a rod ($n=800${, here the energy is computed assuming quadratic functionals, a suitable representation for an inextensible rod}) simulated assuming no dissipation $\gamma=0$ (red line) versus the theoretical total energy $E_F=(M_h\Phi+T_hD)/2$ (black dashed line). (e) Equilibrium rod configuration $\mathbf{r}^n_{eq}$ numerically obtained given the discretization $n=800$, and assuming dissipation. For all studies, unless specified otherwise, we used the following settings: length $L=100$~m, twist $\Phi=27\cdot 2\pi$, slack $D=3$~m, linear mass density $\rho=1$~kg/m, bending stiffness $\alpha=1.345$~Nm$^2$, twisting stiffness $\beta=0.789$~Nm$^2$, shear/stretch matrix $\mathbf{S}=10^5\cdot\pmb{1}$~N, {bend/twist} matrix $\mathbf{B}=\text{diag}(\alpha, \alpha, \beta)$~Nm$^2$, dissipation constant $\gamma=10^{-2}$~kg/(ms), radius $r=0.35$~m, twisting time $T_{\text{twist}}=500$~s, relaxation time $T_{\text{relax}}=10^4$~s.}}
\label{fig:helicalBuckling_spaceconv}
\end{center}
\end{figure}

We validate our discrete derivative operators beyond the onset of instability (see Euler and Mitchell buckling tests in the Appendix) for a long straight, isotropic, inextensible, and unshearable rod undergoing bending and twisting. The filament is characterized by the length $L$ and by the bending and twist stiffnesses $\alpha$ and $\beta$. The clamped ends of the rod are pulled together in the axial direction $\mathbf{k}$ with a slack $D/2$ and simultaneously twisted by the angle $\Phi/2$, as illustrated in Fig.~\ref{fig:helicalBuckling_spaceconv}a. Under these conditions the filament buckles into a localized helical shape (Fig.~\ref{fig:helicalBuckling_spaceconv}e).

The nonlinear equilibrium configuration $\mathbf{r}_{eq}$ of the rod can be analitycally determined \cite{Coyne:1990,vanDerHeijden:2000,Neukirch:2002,vanDerHeijden:2003} in terms of the total applied slack $D$ and twist $\Phi$. We denote the magnitude of the twisting torque and tension acting on both ends and projected on $\mathbf{k}$ by $M_h$ and $T_h$, respectively. Their normalized counterparts $m_h=M_h L/(2\pi\alpha)$ and $t_h=T_hL^2/(4\pi^2\alpha)$ can be computed via the `semi-finite' correction approach \cite{Neukirch:2002} by solving the system
\[
\begin{dcases}
    \frac{D}{L}=\sqrt{\frac{4}{\pi^2t_h}\left(1-\frac{m_h^2}{4t_h}\right)},\\
    \Phi = \frac{2\pi m_h}{\beta/\alpha}+4\arccos\left(\frac{m_h}{2\sqrt{t_h}}\right).
\end{dcases}
\]
Then, the analytical form of $\mathbf{r}_{eq}$ can be expressed \cite{vanDerHeijden:2003} as 
\begin{eqnarray}
\mathbf{r}_{eq}&=&L\left[\frac{1}{2\pi t_{h}}\sqrt{4t_{h}-m^2_{h}} \text{ sech}\left(\pi \bar{s} \sqrt{4t_h-m^2_h}\right)\sin(m_h\pi \bar{s})\right]\mathbf{i}\nonumber\\
&&-L\left[\frac{1}{2\pi t_h}\sqrt{4t_h-m^2_h} \text{ sech}\left(\pi \bar{s} \sqrt{4t_h-m^2_h}\right)\cos(m_h\pi \bar{s})\right]\mathbf{j}\label{eq:helicalAnal}\\
&&+L\left[\bar{s}-\frac{1}{2\pi t_h}\sqrt{4t_h-m^2_h} \text{ tanh}\left(\pi \bar{s} \sqrt{4t_h-m^2_h}\right)\right]\mathbf{k}\nonumber,
\end{eqnarray}
where $\bar{s}=s/L-0.5$ is the normalized arc-length $-0.5\le\bar{s}\le0.5$. Here we make use of Eq.~(\ref{eq:helicalAnal}) to investigate the convergence properties of our solver in the limit of refinement. To compare analytical and numerical solutions, a metric invariant to rotations about $\mathbf{k}$ is necessary. Following Bergou et al.\cite{Bergou:2008}, we rely on the definition of the envelope $\varphi$
\begin{equation}
\varphi = \frac{\cos\theta-\cos \theta_{max}}{1-\cos \theta_{max}},~~~~~~\theta = \arccos(\mathbf{t}\cdot \mathbf{k})\\
\end{equation}
where $\theta$ is the angular deviation of the tangent $\mathbf{t}$ from the axial direction $\mathbf{k}$, and $\theta_{max}$ is the corresponding maximum value along the filament. The envelope $\varphi$ relative to the analytical solution of Eq.~(\ref{eq:helicalAnal}), and $\varphi^n$ relative to a numerical model of $n$ discretization elements can be {estimated via} finite differences. This allows us to determine the convergence order of the solver by means of the norms $L^1(\epsilon)$, $L^2(\epsilon)$ and $L^{\infty}(\epsilon)$ of the error $\epsilon=\|\varphi-\varphi^n\|$.

We simulate the problem illustrated in Fig.~\ref{fig:helicalBuckling_spaceconv} at different space-time resolutions. The straight rod originally at rest is twisted and compressed at a constant rate during the period $T_{\text{twist}}$. Subsequently, the ends of the rod are held in their final configurations for the period $T_{\text{relax}}$ to allow the internal energy to dissipate (according to the model of Section~\ref{sec:dissipation}) until the steady state is reached. Simulations are carried out progressively refining the spatial discretization $\delta l = L/n$ by varying $n=100-3200$ and the time discretization $\delta t$ is kept proportional to $\delta l$, as reported in Fig.~\ref{fig:helicalBuckling_spaceconv}.

As can be seen in Fig.~(\ref{fig:helicalBuckling_spaceconv})b,c,e the numerical solutions converge to the analytical one with second order in time and space, consistent with our spatial and temporal discretization schemes. Moreover, to further validate the energy conserving properties of the solver, we turn off the internal dissipation (Fig.~\ref{fig:helicalBuckling_spaceconv}d) and observe that the total energy of the filament $E_F$ is constant after $T_{\text{twist}}$ and matches its theoretical value $E_F=(M_h\Phi+T_hD)/2$.

\subsubsection{Vertical oscillations under gravity}

\begin{figure}
\begin{center}
\includegraphics[width=\textwidth]{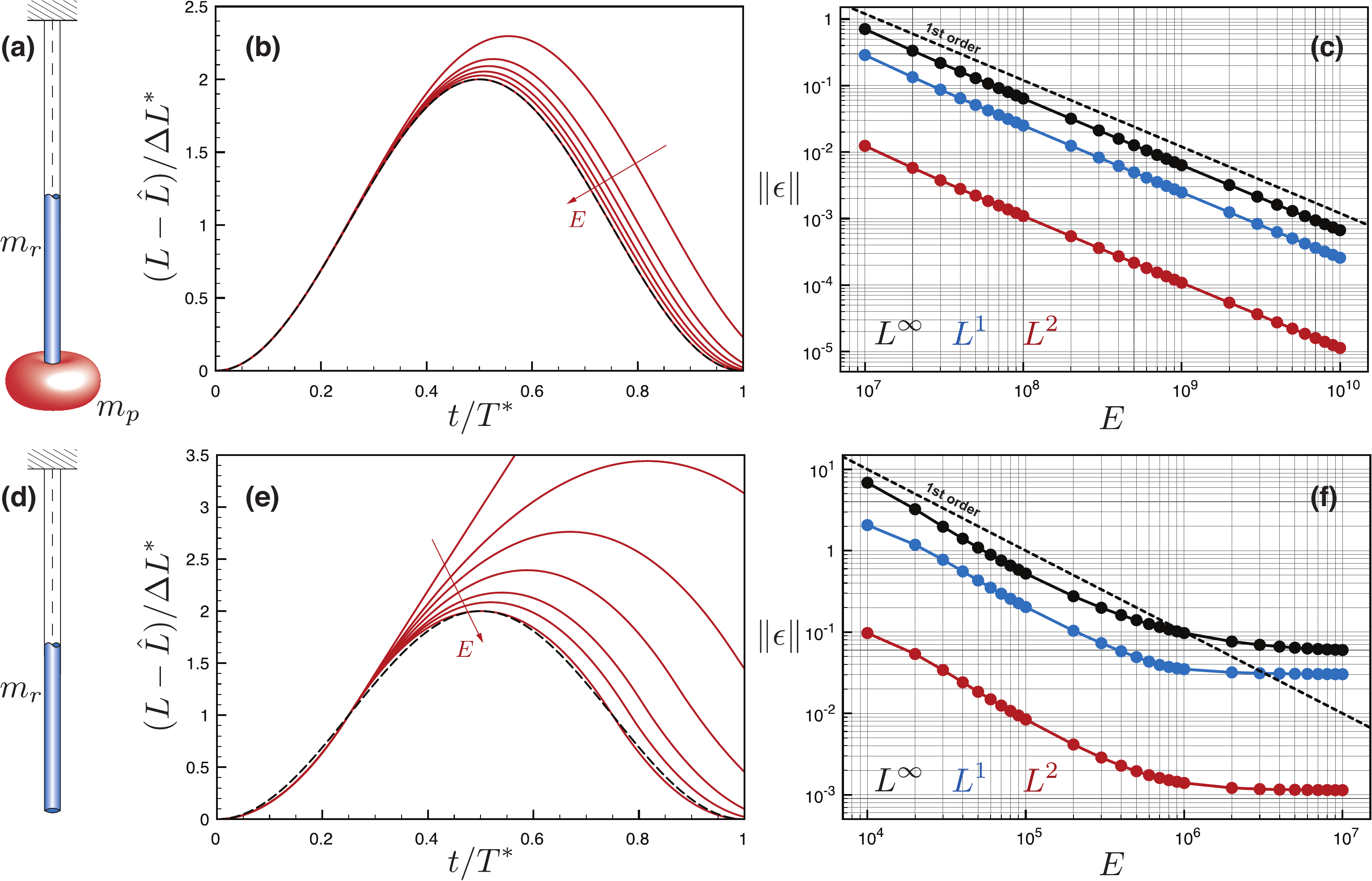}
\caption{\scriptsize{\textbf{Vertical oscillation under gravity.} (a,d) We consider a vertical rod of mass $m_r$ clamped at the top and with a mass $m_p$ attached to the free end. Assuming that the rod is stiff enough (i.e. $k\simeq\hat{A}E=\text{const}$), it oscillates due to gravity around the equilibrium position $\hat{L}+\Delta L^*$, where $\Delta L^*=g(m_p+m_r/2)/k$ with a period $T^*=2\pi\sqrt{(m_p+m_r/\xi)/k}$ with $\xi\simeq3$ for $m_p\gg m_r$, and $\xi\simeq\pi^2/4$ for $m_p\ll m_r$. Therefore, the rod oscillates according to $L(t)=\hat{L} + [1 + \sin(2\pi t/T^* -\pi/2)]\Delta L^*$. (a-b) Case $m_p\gg m_r$ with $m_p=100$~kg and $m_r=1$~kg. (b) By increasing the stiffness $E=10^7$, 2$\cdot$$10^7$, 3$\cdot$$10^7$, 5$\cdot$$10^7$, $10^8$, $10^{10}$~Pa, the simulated oscillations (red lines) approach the analytical solution (dashed black line). (c) Convergence to the analytical solution in the norms $L^{\infty}(\epsilon)$ (black), $L^1(\epsilon)$ (blue) and $L^2(\epsilon)$ (red) with $\epsilon=\|L(t)-L^E(t)\|$, where $L^E$ is the length numerically obtained as a function of $E$. (c-d) Case $m_p\ll m_r$ with $m_p=0$~kg and $m_r=1$~kg. (e) By increasing the stiffness $E=10^4$, 2$\cdot$$10^4$, 3$\cdot$$10^4$, 5$\cdot$$10^4$, $10^5$, 2$\cdot$$10^5$, $10^9$~Pa, the simulated oscillations approach the analytical solution. (f) Convergence to the analytical solution in the norms $L^{\infty}(\epsilon)$, $L^1(\epsilon)$ and $L^2(\epsilon)$ as a function of $E$. For all studies, we used the following settings: gravity $g=9.81$~m/s$^2$, rod density $\rho=10^3$~kg/m$^3$, shear modulus $G=2E/3$~Pa, shear/stretch matrix $\hat{\mathbf{S}}=\text{diag}(4G\hat{A}/3, 4G\hat{A}/3, E\hat{A})$~N, {bend/twist} matrix $\hat{\mathbf{B}}=\text{diag}(E\hat{I}_1, E\hat{I}_2, G\hat{I}_3)$~Nm$^2$, rest length $\hat{L}=1$~m, rest cross sectional area $\hat{A}=~m_r/(\hat{L}\rho)$~m$^2$, number of discretization elements $n=100$, timestep $\delta t = T^*/10^6$, dissipation constant $\gamma=0$.}}
\label{fig:oscillationGravity}
\end{center}
\end{figure}

We consider a system in which a rod hanging from one end and subject to gravity $g$ oscillates due to a mass $m_p$ suspended at the other end, and due to its own mass $m_r$, as depicted in Fig.~\ref{fig:oscillationGravity}a,d. This system is analogous to a mass-spring oscillator. The static solution is then obtained by integrating the infinitesimal elongations along the spring due to the local load \cite{Galloni:1979}, yielding the total equilibrium extension
\begin{equation}
\Delta L^*=\frac{gm_{eq}}{k}=\frac{g(m_p+m_r/\xi)}{k},
\end{equation}
where $k$ is the spring constant, $\xi=2$ is a constant factor, and $m_{eq}=m_p+m_r/\xi$ is the equivalent mass. Thus, the final equilibrium length of the rod reads $L=\hat{L}+\Delta L^*$, with $\hat{L}$ being the rest unstretched length.

The dynamic solution is instead characterized by oscillations of period $T^*$ and by {the} time varying length $L(t)$ of the spring
\begin{equation}
T^*=2\pi\sqrt{(m_p+m_r/\xi)/k},~~~~~L=\hat{L} + [1 + \sin(2\pi t/T^* -\pi/2)]\Delta L^*.
\end{equation}
In this case, unlike the static solution, the factor $\xi$ depends on the ratio $m_r/m_p$. In fact it can be shown \cite{Galloni:1979} that $\xi \simeq 3$ for $m_r/m_p\rightarrow 0$, and $\xi \simeq \pi^2/4$ for $m_r/m_p\rightarrow \infty$.

The analytical results rely on the assumption of $k$ being constant in space and time, given a fixed ratio $m_r/m_p$. However, this condition is not met here since $k(s,t)=EA(s,t)$ is a function of space and time, due to dilatation and mass conservation. Nevertheless, as the Young's modulus $E\rightarrow\infty$, that is as a soft filament becomes stiff, the constant $k\rightarrow E\hat{A}$ and our rod model must recover the behavior of the mass-spring oscillator. Indeed, Fig.~\ref{fig:oscillationGravity}b,c,e,f shows how the proposed numerical method converges to the analytical oscillation period $T^*$ and normalized longitudinal displacement $(L-\hat{L})/\Delta L^*$ as $E$ increases.

\subsubsection{Cantilever beam}
\label{sec:timo}
\begin{figure}
\begin{center}
\includegraphics[width=\textwidth]{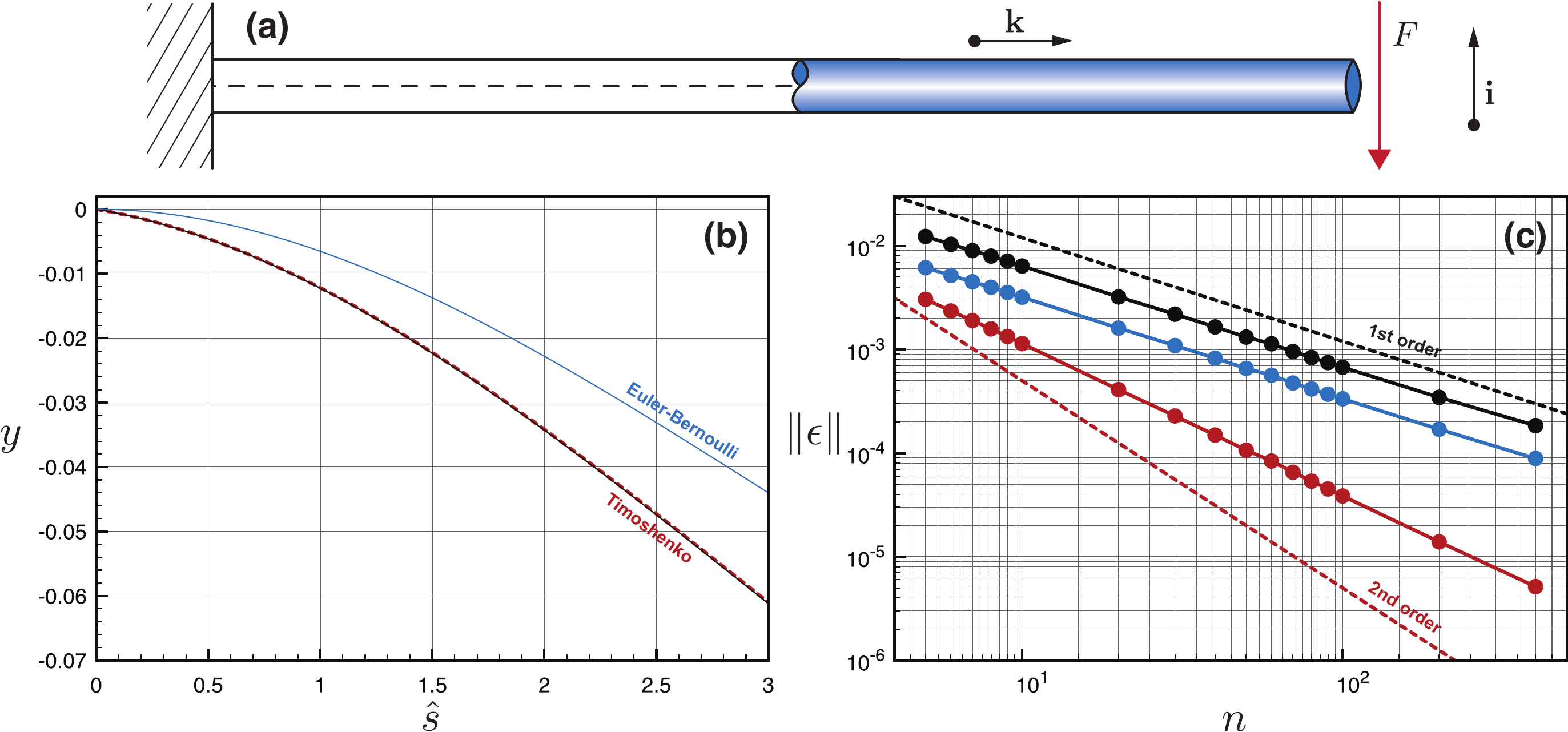}
\caption{\scriptsize{\textbf{Time-space convergence study for a cantilever beam.} (a) We consider the static solution of a beam clamped at one end $\hat{s}=0$ and subject to the downward force $F$ at the free end $\hat{s}=\hat{L}$. (b) Comparison between the Timoshenko analytical $y$ (black lines) and numerical $y^n$ (with $n=400$, red dashed lines) vertical displacements with respect to the initial rod configuration. As a reference we report in blue the corresponding Euler-Bernoulli solution. (c) Norms $L^{\infty}(\epsilon)$ (black), $L^1(\epsilon)$ (blue) and $L^2(\epsilon)$ (red) of the error $\epsilon=\|y-y^n\|$ at different levels of time-space resolution are plotted against the number of discretization elements $n$. Here, the time discretization $\delta t$ is slaved by the spatial discretization $n$ according to $\delta t = 10^{-2} \delta l$ seconds. For all studies, we used the following settings: rod density $\rho=5000$~kg/m$^3$, Young's modulus $E=10^6$~Pa, shear modulus $G=10^4$~Pa, shear/stretch matrix $\hat{\mathbf{S}}=\text{diag}(4G\hat{A}/3, 4G\hat{A}/3, E\hat{A})$~N, {bend/twist} matrix $\hat{\mathbf{B}}=\text{diag}(E\hat{I}_1, E\hat{I}_2, G\hat{I}_3)$~Nm$^2$, downward force $F=15$~N, rest length $\hat{L}=3$~m, rest radius $\hat{r}=0.25$~m, dissipation constant $\gamma=10^{-1}$~kg/(ms), simulation time $T_{\text{sim}}$=$5000$~s.}}
\label{fig:timoshenko_conv}
\end{center}
\end{figure}

We consider now the effect of {bend} and shear simultaneously by validating our numerical methods against the Timoshenko cantilever of Fig.~\ref{fig:timoshenko_conv}a. Timoshenko's model accounts for bending elasticity, rotary inertia and shear deformations, building on classical beam theories by Rayleigh (bending elasticity and rotary inertia) and Euler-Bernoulli (bending elasticity only). The model captures the behavior of short or composite beams in which shear deformations effectively lower the stiffness of the rod \cite{Rao:1995,Gere:2001}.

We consider a beam clamped at one end $\hat{s}=0$ and subject to the downward force $F$ at the free end $\hat{s}=\hat{L}$, as illustrated in Fig.~\ref{fig:timoshenko_conv}a. The static solution for the displacement $y$ along the vertical direction $\mathbf{i}$ of the rod can then be analytically expressed  as
\begin{equation}
y=-\frac{F}{\alpha_c\hat{A}G}\hat{s} - \frac{F\hat{L}}{2E\hat{I}_1}\hat{s}^2 + \frac{F}{6E\hat{I}_1}\hat{s}^3,
\label{eq:displacementTimo}
\end{equation}
where $\hat{L}$ is the length of the rod, $\hat{A}$ is the constant cross sectional area, $\hat{I}_1$ is the area second moment of inertia about the axis $\mathbf{j}=\mathbf{k}\times\mathbf{i}$, $E$ and $G$ are the Young's and shear moduli, and $\alpha_c=4/3$ is the Timoshenko shear factor for circular sections and accounts for the fact that the shear stress varies over the section \cite{Gere:2001}. Furthermore, the Timoshenko (as well as Rayleigh and Euler-Bernoulli) theory relies on the assumption of small deflections, so that the horizontal coordinate $x$ along the direction $\mathbf{k}$ can be approximated by the arc-length $\hat{s}$ (Fig.~\ref{fig:timoshenko_conv}a and \textit{Appendix} for further details and derivation), hence the use of $\hat{s}$ in the above equation .

If the shear modulus $G$ approaches infinity or if the ratio $E\hat{I}_1/(\alpha_c\hat{L}^2\hat{A}G)\gg 1$, then the Timoshenko model in the static case reduces to the Euler-Bernoulli approximation, yielding
\begin{equation}
y= - \frac{F\hat{L}}{2E\hat{I}_1}\hat{s}^2 + \frac{F}{6E\hat{I}_1}\hat{s}^3,
\end{equation}
{as} the shear term of Eq.~(\ref{eq:displacementTimo}) becomes negligible.

We compare our numerical model with these results by carrying out simulations of the cantilever beam of Fig.~\ref{fig:timoshenko_conv}a in the time-space limit of refinement. As can be noticed in Fig.~\ref{fig:timoshenko_conv}b the discrete solution recovers the Timoshenko one. Therefore, the solver correctly captures the role of shear that reduces the effective stiffness relative to the Euler-Bernoulli solution. Moreover, our approach is shown to converge to the analytical solution in all the norms $L^\infty(\epsilon)$, $L^1(\epsilon)$, $L^2(\epsilon)$ of the error $\epsilon=\|y-y^n\|$, where $y^n$ is the vertical displacement numerically obtained in the refinement limit.

We note that the norms $L^\infty(\epsilon)$ and $L^1(\epsilon)$ exhibit first order convergence, while $L^2(\epsilon)$ decays with a slope between first and second order. We attribute these results to the fact that while the Timoshenko solution does not consider axial extension or tension, it does rely on the assumption of small deflections ($\hat{s}=x$), therefore effectively producing a dilatation of the rod. On the contrary, our solver does not assume small deflections and does not neglect axial extension, since the third entry of the matrix $\mathbf{B}$ has the finite value $E\hat{A}$ (see Fig.~\ref{fig:timoshenko_conv} for details). This discrepancy is here empirically observed to {decrease} the convergence order. 

{These studies, together with the ones reported in the Appendix, complete the validation of the proposed numerical scheme and demonstrate the accuracy of our methodology in simulating soft filaments in simple settings.}

\section{Including interactions and activity: solid and liquid friction, {contact and}  muscular effects}
\label{sec:additionalphysics}

{Motivated by the advancements in the field of soft robotics  \cite{Kim:2013,Haines:2014,Park:2016}, we wish to develop a robust and accurate framework for the characterization and computational design of soft slender structures interacting with complex environments. To this end, we expand the range of applications of our formalism by including} additional physical effects, from viscous hydrodynamic forces in the slender-body limit and surface solid friction to self-contact and active muscular activity. {As a general strategy, all new  \textit{external} physical  interactions are accounted for by lumping their contributions into the external forces and couples $\mathbf{F}$ and $\mathbf{C}_{\mathcal{L}}$ on the right hand side of the linear and angular momentum {balance Eqs.}~(\ref{eq:linmomentfinal}, \ref{eq:angmomentfinal}). On the other hand, all new  \textit{internal} physical and biophysical effects are captured by adding their {contributions} directly to the internal force $\mathbf{n}_{\mathcal{L}}$ and torque $\boldsymbol{\tau}_{\mathcal{L}}$ resultants before {integrating} Eqs.~(\ref{eq:linmomentfinal}, \ref{eq:angmomentfinal}).}

\subsubsection{Dissipation} 
 
\label{sec:dissipation}
Real materials are subject to internal friction and viscoelastic losses, which can be modeled by modifying the constitutive relations so that the internal toques $\boldsymbol{\tau}_{\mathcal{L}}(\boldsymbol{\kappa}_{\mathcal{L}})$ and forces $\mathbf{n}_{\mathcal{L}}(\boldsymbol{\sigma}_{\mathcal{L}})$ of Eqs.~(\ref{eq:constitutivelaws}) become functions of both strain and rate of strain, i.e. $\boldsymbol{\tau}_{\mathcal{L}}(\boldsymbol{\kappa}_{\mathcal{L}}, \partial_t\boldsymbol{\kappa}_{\mathcal{L}})$ and $\mathbf{n}_{\mathcal{L}}(\boldsymbol{\sigma}_{\mathcal{L}}, \partial_t \boldsymbol{\sigma}_{\mathcal{L}})$. Keeping track of the strain rates increases computational costs and the memory footprint of the solver. However, for the purpose of purely dissipating energy, a simple alternative option is to employ Rayleigh potentials \cite{Torby:1984,Audoly:2013}. In this case viscous forces $\mathbf{f}_v$ and torques $\mathbf{c}^v_{\mathcal{L}}$ per unit length are directly computed as linear functions of linear and angular velocities through the constant translational $\gamma_t$ and rotational $\gamma_r$ internal friction coefficients, so that
\begin{eqnarray}
\mathbf{f}_v&=&-\gamma_t\mathbf{v}\label{eq:dissipationvel},\\
\mathbf{c}^v_{\mathcal{L}}&=&-\gamma_r\boldsymbol{\omega}_{\mathcal{L}}.\label{eq:dissipationangvel}
\end{eqnarray}
This approach does not model the physical nature of viscoelastic phenomena, although it does dissipate energy, effectively mimicking overall material friction effects. In the context of our numerical investigations, we did not observe any appreciable difference between the two outlined methods, so that, for the sake of simplicity and computational efficiency, we opted for the second one. Throughout the remainder of the text we will then employ Eqs.~(\ref{eq:dissipationvel}, \ref{eq:dissipationangvel}) with a single dissipation constant $\gamma$, therefore assuming $\gamma_t=\gamma_r$. 

\subsubsection{Muscular activity}
\label{sec:muscularActivity}
To study limbless biolocomotion on solid substrates and in fluids, we allow our soft filaments to be active, by generating internal forces and torques corresponding to coordinated muscular activity driven, for example, by a central pattern generator \cite{Altringham:1990,Gazzola:2015}.

\begin{figure}
\begin{center}
\includegraphics[width=0.90\textwidth]{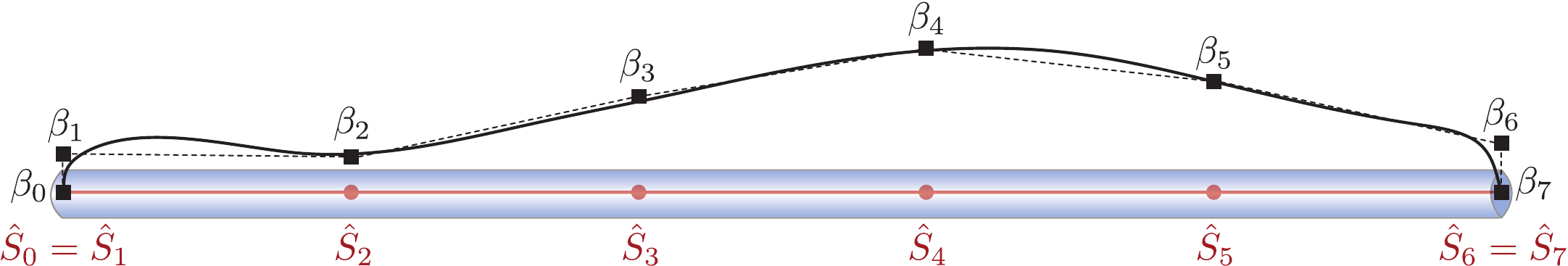}
\caption{\scriptsize{\textbf{Muscular activity.} Example of muscular activity amplitude profile (solid black line) described by cubic B-spline through $N_m=8$ control points $(\hat{S}_i,\beta_i)$ with $i=0,\dots,N_m-1$. The control points are located along the filament at the positions $\hat{S}_i$, and are associated with the amplitude values $\beta_i$. The first and last control points are fixed so that $(\hat{S}_0,\beta_0)=(0,0)$ and $(\hat{S}_{N_m-1},\beta_{N_m-1})=(\hat{L},0)$, therefore assuming the ends of the deforming body to be free.}}
\label{fig:muscularActivity}
\end{center}
\end{figure}

Following the approach detailed in \cite{Gazzola:2012,Rees:2013}, we express the muscular activity magnitude $A_m$ as a traveling wave propagating head to tail along the filament 
\begin{equation}
A_m=\beta_m(\hat{s})\cdot\sin\left(\frac{2\pi}{T_m} t + \frac{2\pi}{\lambda_m} \hat{s}+\phi_m\right),
\end{equation}
where $\phi_m$ is the phase, $t$ is time, $T_m$ and $\lambda_m$ are, respectively, the activation period and wavelength. The amplitude of the traveling wave is represented by the cubic B-spline $\beta(\hat{s})$ characterized by $N_m$ control points $(\hat{S}_i,\beta_i)$ with $i=0,\dots,N_m-1$, as illustrated in Fig.~\ref{fig:muscularActivity}. The first and last control points are fixed so that $(\hat{S}_0,\beta_0)=(0,0)$ and $(\hat{S}_{N_m-1},\beta_{N_m-1})=(\hat{L},0)$, therefore assuming the ends of the deforming body to be free. One of the main advantages of the proposed parametrization is that it encompasses a large variety of patterns with a relatively small number of parameters \cite{Rees:2013}.

A given activation mode can be achieved by multiplying the scalar amplitude $A_m$ with the appropriate director. For example, if we wish to study earthworm-like locomotion we may employ a wave of longitudinal dilatation and compression forces, so that
\begin{equation}
\mathbf{n}^m_{\mathcal{L}}  =  \mathbf{Q}(A_m\mathbf{d}_3).
\end{equation}
Similarly, if we wish to investigate a slithering snake characterized by a planar kinematic wave, we may consider a torque activation of the form
\begin{equation}
\boldsymbol{\tau}^m_{\mathcal{L}} = \mathbf{Q}(A_m\mathbf{d}_1),
\end{equation}
assuming $\mathbf{d}_2$ and $\mathbf{d}_3$ to be coplanar to the ground. {These two contributions are directly added to the internal force $\mathbf{n}_{\mathcal{L}}$ and torque $\boldsymbol{\tau}_{\mathcal{L}}$ resultants.}

In the most general case, all deformation modes can be excited by enabling force and torque muscular activity along all directors $\mathbf{d}_1$, $\mathbf{d}_2$ and $\mathbf{d}_3$, providing great flexibility in terms of possible gaits.

\subsubsection{Self-contact} 
{To prevent the filament from passing through itself, we need to account for self-contact. As a general strategy, we avoid enforcing the presence of boundaries via Lagrangian constraints as their formulation may be cumbersome \cite{Riewe:1996}, impairing the modularity of the numerical solver. We instead resort to calculating forces and torques directly and replacing hard constraints with `soft' displacement-force relations.}

{Our self-contact model introduces additional} forces $\mathbf{F}_{sc}$ acting between the discrete elements in contact. To determine whether any two cylindrical elements are in contact, we calculate the minimum distance $d_{min}^{ij}$ between edges $i,j$ by parameterizing their centerlines $c_i(h) = s_i+h(s_{i+1}-s_i)$ so that
\begin{equation}
d_{min}^{ij}= \max\limits_{h_1,h_2\in[0,1]}||c_i(h_1) - c_j(h_2)||.
\end{equation}
If $d_{min}^{ij}$ is smaller than the sum of the radii of the two cylinders, then they are considered to be in contact and penalty forces are applied to each element as a function of the scalar overlap $\epsilon_{ij} = (r_i + r_j - d_{min}^{ij})$, where $r_i$ and $r_j$ are the radii of edges $i$ and $j$. If $\epsilon_{ij}$ is smaller than zero, then the two edges are not in contact and no penalty is applied. Denoting as $\mathbf{d}^{ij}_{min}$ the unit vector pointing from closest point on edge $i$ to the closest point on edge $j$, the self-contact repulsion force is given by
\begin{equation}
\mathbf{F}_{sc}= H(\epsilon_{ij})\cdot\left[-k_{sc}\epsilon_{ij}-\gamma_{sc}(\mathbf{v}_i-\mathbf{v}_j)\cdot\mathbf{d}_{min}^{ij}\right]\mathbf{d}_{min}^{ij}
\end{equation}
where $H(\epsilon_{ij})$ denotes the Heaviside function and ensures that a repulsion force is produced only in case of contact ($\epsilon_{ij}\ge0$). The first term within the square brackets expresses the linear response to the interpenetration distance as modulated by the stiffness $k_{sc}$, while the second damping term models contact dissipation and is proportional to the coefficient $\gamma_{sc}$ and the interpenetration velocity $\mathbf{v}_i-\mathbf{v}_j$.

\subsubsection{Contact with solid boundaries}
\label{sec:solidBoundaries}
\begin{figure}
\begin{center}
\includegraphics[width=0.40\textwidth]{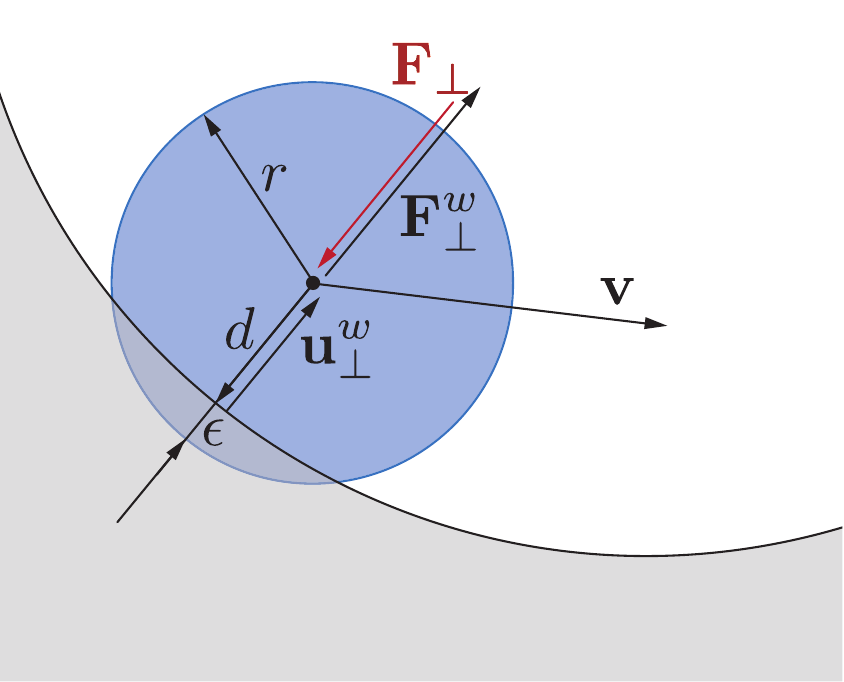}
\caption{\scriptsize{\textbf{Contact model with solid boundaries.} Obstacles and surfaces (gray) are modeled as soft boundaries allowing for the interpenetration $\epsilon=r-d$ with the elements of the filament (blue) characterized by radius $r$ and distance $d$ from the substrate. The surface normal $\mathbf{u}^w_{\perp}$ determines the direction of the wall's response $\mathbf{F}^w_{\perp}$ to contact. We note that $\mathbf{F}^w_{\perp}$ balances the sum of all forces $\mathbf{F}_{\perp}$ that push the rod against the wall, and is complemented by other two components which allow to amend to possible interpenetration due to numerics. These components are an elastic one ($k_w\epsilon$) and a dissipative one ($\gamma_w\mathbf{v}\cdot \mathbf{u}^w_{\perp}$), where $k_w$ and $\gamma_w$ are, respectively, the wall stiffness and dissipation coefficients.}}
\label{fig:walls}
\end{center}
\end{figure}

{In order {to} investigate scenarios in which filaments interact with the surrounding environment, we must also account for solid boundaries. By implementing the same approach outlined in the previous section, obstacles and surfaces are  modeled as soft boundaries} allowing for interpenetration with the elements of the rod (Fig.~\ref{fig:walls}). The wall response $\mathbf{F}^w_{\perp}$ balances the sum of all forces $\mathbf{F}_{\perp}$ that push the rod against the wall, and is complemented by other two components which help prevent possible interpenetration due to numerics. The interpenetration distance $\epsilon$ triggers a normal elastic response proportional to the stiffness of the wall while a dissipative term related to the normal velocity component of the filament with respect to the substrate accounts for a   damping force, so that the overall wall response reads
\begin{equation}
\mathbf{F}^w_{\perp}= H(\epsilon)\cdot(-\mathbf{F}_{\perp} + k_w\epsilon-\gamma_w\mathbf{v}\cdot \mathbf{u}^w_{\perp})\mathbf{u}^w_{\perp}
\end{equation}
where $H(\epsilon)$ denotes the Heaviside function and ensures that a wall force is produced only in case of contact ($\epsilon\ge0$). Here $\mathbf{u}^w_{\perp}$ is the boundary outward normal (evaluated at the contact point, that is the contact location for which the normal passes through the center of mass of the element), and $k_w$ and $\gamma_w$ are, respectively, the wall stiffness and dissipation coefficients.

\subsubsection{Isotropic and anisotropic surface friction} 
\label{sec:friction}
Solid boundaries also affect the dynamics of the filament through surface friction, a complex physical phenomenon in which a range of factors are involved, from roughness and plasticity of the surfaces in contact to the kinematic initial conditions and geometric setup. Here, we adopt the Amonton-Coulomb model, the simplest of friction models.

This model relates the normal force pushing a body onto a substrate to the friction force through the kinetic $\mu_k$ and static $\mu_s$ friction coefficients, depending on whether the contact surfaces are in relative motion or not.

\begin{figure}
\begin{center}
\includegraphics[width=\textwidth]{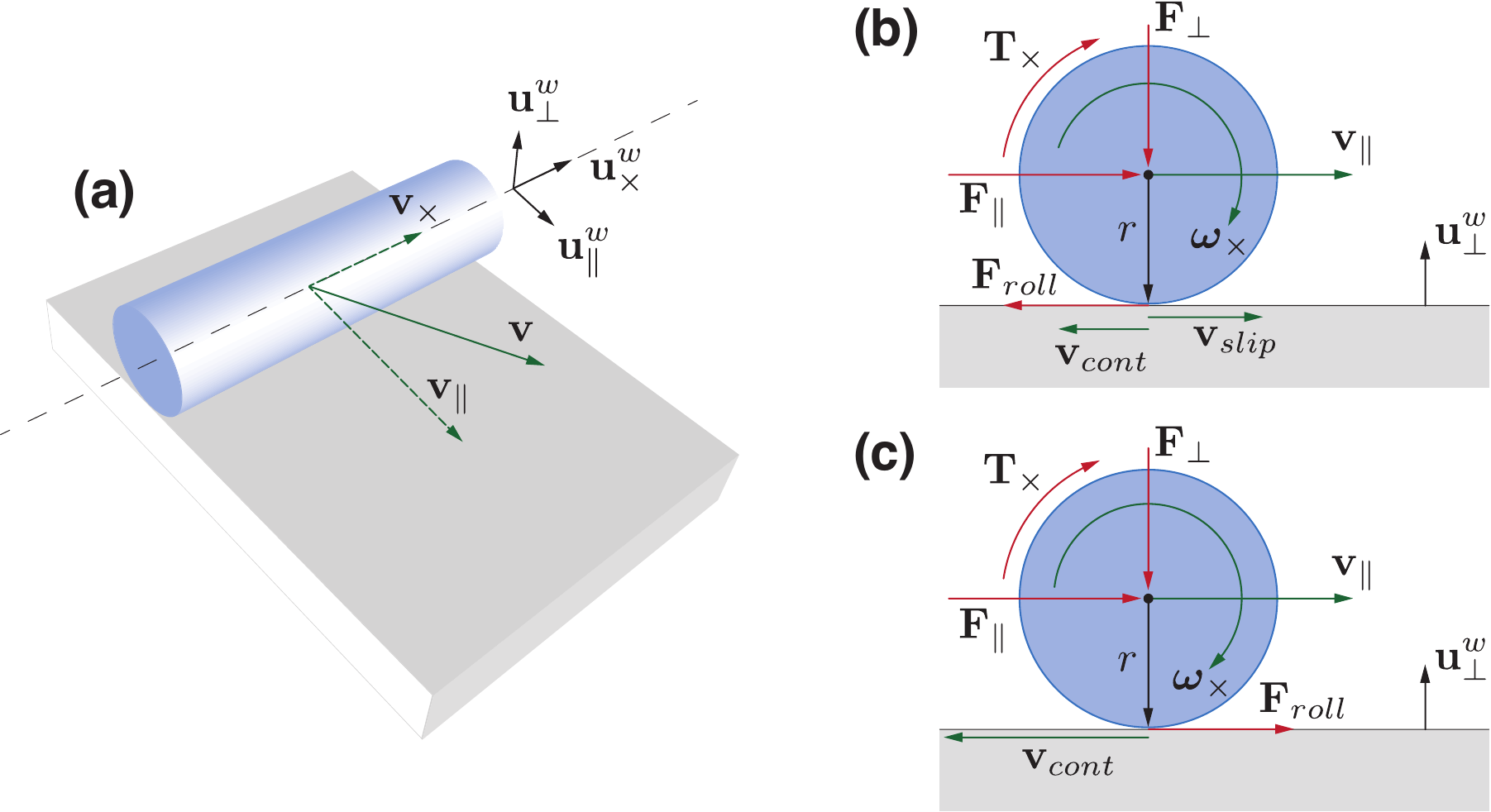}
\caption{\scriptsize{\textbf{Surface friction.} (a) The forces produced by friction effects between an element of the rod and the substrate are naturally decomposed into a lateral component in the direction $\mathbf{u}^w_{\parallel}= \mathbf{t}\times \mathbf{u}^w_{\perp}$ and a longitudinal one in the direction $\mathbf{u}^w_{\times}= \mathbf{u}^w_{\perp}\times \mathbf{u}^w_{\parallel}$. We note that in general $\mathbf{u}^w_{\times}\neq \mathbf{t}$. The notation $\mathbf{x}_{\perp}$, $\mathbf{x}_{\parallel}$, $\mathbf{x}_{\times}$ denotes the projection of the vector $\mathbf{x}$ in the directions $\mathbf{u}^w_{\perp}$, $\mathbf{u}^w_{\parallel}$, $\mathbf{u}^w_{\times}$. (b,c) Kinematic and dynamic quantities at play at any cross section in case of rolling and slipping (b) and pure rolling (c) motion. Red arrows correspond to forces and torques, green arrows correspond to velocities, and black arrows correspond to geometric quantities.}}
\label{fig:frictionModel}
\end{center}
\end{figure}

Despite the simplicity of the model, its formulation and implementation may not {necessarily be} straightforward, especially in the case of rolling motions. Given the cylindrical geometry of our filaments, the effect of surface friction can be decomposed into a longitudinal component associated with purely translational displacements, and a lateral component associated with both translational and rotational motions (Fig.~\ref{fig:frictionModel}). We use the notation $\mathbf{x}_{\perp}$, $\mathbf{x}_{\parallel}$, $\mathbf{x}_{\times}$ to denote the projection of the vector $\mathbf{x}$ in the directions $\mathbf{u}^w_{\perp}$, $\mathbf{u}^w_{\parallel}$, $\mathbf{u}^w_{\times}$, as illustrated in Fig.~\ref{fig:frictionModel}.

The longitudinal friction force $\mathbf{F}_{long}$ is opposite to either the resultant of all forces $\mathbf{F}_{\times}$ acting on an element (static case) or to the translational velocity $\mathbf{v}_{\times}$ (kinetic case) along the direction $\mathbf{u}^w_{\times}$ (Fig.~\ref{fig:frictionModel}). The Amonton-Coulomb model then reads
\[
\mathbf{F}_{long}= 
\begin{dcases}
    -\max(|\mathbf{F}_{\times}|,\mu_s |\mathbf{F}_{\perp}|)\cdot \frac{\mathbf{F}_{\times}}{|\mathbf{F}_{\times}|}& \text{if } |\mathbf{v}_{\times}|\le v_{\epsilon}\\
    -\mu_k |\mathbf{F}_{\perp}| \cdot \frac{\mathbf{v}_{\times}}{|\mathbf{v}_{\times}|}              & \text{if }  |\mathbf{v}_{\times}| > v_{\epsilon}
\end{dcases},
\]
where $v_{\epsilon}\rightarrow0$ is the absolute velocity threshold value employed to distinguish between static ($|\mathbf{v}_{\times}|\le v_{\epsilon}$) and kinetic ($|\mathbf{v}_{\times}| > v_{\epsilon}$) case. We define $v_{\epsilon}$ in a limit form to accommodate the fact that  inequalities are numerically evaluated up to a small threshold value. The static friction force is always equal and opposite to $\mathbf{F}_{\times}$ up to a maximum value proportional to the normal force $|\mathbf{F}_{\perp}|$ though the coefficient $\mu_s$. The kinetic friction force is instead opposite to the translational velocity $\mathbf{v}_{\times}$, but does not depend on its actual magnitude {and} is proportional to $|\mathbf{F}_{\perp}|$ via $\mu_k$. In general $\mu_s>\mu_k$, so that it is harder to set a body into motion from rest than {to drag} it.

The lateral displacement of a filament in the direction $\mathbf{u}^w_{\parallel}=\mathbf{u}^w_{\times}\times \mathbf{u}^w_{\perp}$ is associated with both translational ($\mathbf{v}_{\parallel}$) and rotational ($\boldsymbol{\omega}_{\times}=\omega_{\times}\mathbf{u}^w_{\times}$) motions, as illustrated in Fig.~\ref{fig:frictionModel}b,c. In this case the distinction between static and kinetic friction does not depend on $\mathbf{v}_{\parallel}$, but on the relative velocity (also referred to as slip velocity) between the rod and the substrate
\begin{equation}
\mathbf{v}_{slip} = \mathbf{v}_{\parallel} + \mathbf{v}_{cont},~~~~~\mathbf{v}_{cont}=r\mathbf{u}^w_{\perp}\times\boldsymbol{\omega}_{\times},
\end{equation}
where $\mathbf{v}_{cont}$ is the local velocity of the filament at the contact point with the substrate, due to the axial component of the angular velocity $\boldsymbol{\omega}_{\times}$.

In the static or no-slip scenario ($\mathbf{v}_{slip}=\mathbf{0}$), the linear momentum balance in the direction $\mathbf{u}^w_{\parallel}$, and the angular momentum balance about the axis $\mathbf{u}^w_{\times}$ express a kinematic constraint between the linear acceleration $a\mathbf{u}^w_{\parallel}$ and angular acceleration $\boldsymbol{\omega}_{\times}=(\mathbf{u}^w_{\perp} \times a\mathbf{u}^w_{\parallel})/r$, so that
\begin{eqnarray}
(F_{\parallel} + F_{roll})\mathbf{u}^w_{\parallel} &=& dm\cdot a\mathbf{u}^w_{\parallel}\\
T_{\times}\mathbf{u}^w_{\times}-r\mathbf{u}^w_{\perp}\times F_{roll}\mathbf{u}^w_{\parallel}&=&J\cdot\frac{\mathbf{u}^w_{\perp} \times a\mathbf{u}^w_{\parallel}}{r},
\end{eqnarray}
where $\mathbf{F}_{\parallel}=F_{\parallel}\mathbf{u}^w_{\parallel}$ and $\mathbf{T}_{\times}=T_{\times}\mathbf{u}^w_{\times}$ are the forces and torques acting on the local element, and $\mathbf{F}_{roll}=F_{roll}\mathbf{u}^w_{\parallel}$ is the rolling friction force at the substrate-filament interface necessary to meet the no-slip condition. By recalling that a disk mass second moment of inertia about $\mathbf{u}^w_{\times}$ is $J=r^2dm/2$, the above system can be solved for the unknown $a$ and $F_{roll}$, yielding
\begin{equation}
\mathbf{F}_{roll} = - \frac{rF_{\parallel}-2T_{\times}}{3r}\mathbf{u}^w_{\parallel}.
\label{eq:noslipforcerotation}
\end{equation}

Therefore the lateral friction force $\mathbf{F}_{lat}$ and the associate torque $\mathbf{C}^{lat}_{\mathcal{L}}$ can be finally expressed as 
\[
\mathbf{F}_{lat}= 
\begin{dcases}
    \max(|\mathbf{F}_{roll}|,\mu^r_s |\mathbf{F}_{\perp}|)\cdot \frac{\mathbf{F}_{roll}}{|\mathbf{F}_{roll}|}& \text{if } |\mathbf{v}_{slip}|\le v_{\epsilon}\\
    -\mu^r_k |\mathbf{F}_{\perp}| \cdot \frac{\mathbf{v}_{slip}}{|\mathbf{v}_{slip}|}              & \text{if }  |\mathbf{v}_{slip}| > v_{\epsilon}
\end{dcases},
~~~~~~~~\mathbf{C}^{lat}_{\mathcal{L}}=\mathbf{Q}(\mathbf{F}_{lat}\times r\mathbf{u}_{\perp}^w),
\]
where $\mu_s^r$ and $\mu_k^r$ are, respectively, the rolling static and kinetic friction coefficients.

{So far we have considered} isotropic friction by assuming that the coefficients $\mu_s$ and $\mu_k$ are constant and independent from the direction of the total acting forces (static case) or relative velocities (kinetic case). Nevertheless, frictional forces may be highly anisotropic. For example, the anisotropy caused by the presence of scales on the body of a snake crucially affects gaits and performance \cite{Guo:2008,Hu:2009}.

The Amonton-Coulomb model can be readily extended to account for anisotropic effects by simply assuming the friction coefficients $\mu_s$ and $\mu_k$ to be functions of a given reference direction. The nature of these functions depends on the specific physical problems under investigation. An example of this approach is illustrated in Section \ref{sec:limblessLocomotion} in the context of limbless locomotion. {Isotropic and anisotropic friction validation benchmarks are presented in the Appendix.}

\subsubsection{Hydrodynamics}
\label{sec:hydro}
We also extend our computational framework to address flow-structure interaction problems. In particular we consider the case in which viscous forces dominate over inertial effects, i.e. we consider systems in which the Reynolds number $Re=\rho_f U L/\mu\ll1$ where $\rho_f$ and  $\mu$ are the density and dynamic viscosity of the fluid, and $U$ is the characteristic velocity of the rod. {Under} these conditions, the drag forces exerted by the fluid on our filaments can be determined analytically within the context of slender-body theory \cite{Cox:1970,Lauga:2009}. At leading order resistive {force} line densities scale linearly with the local rod velocities $\mathbf{v}$ according to
\begin{equation}
\mathbf{f}^H = -\frac{4\pi\mu}{\ln(L/r)}\left(\mathbf{I}-\frac{1}{2}\mathbf{t}^T\mathbf{t}\right)\mathbf{v}.
\end{equation}
We note that the matrix $(\mathbf{I}-\frac{1}{2}\mathbf{t}^T\mathbf{t})$ introduces an anisotropic effect for which
\begin{equation}
\mathbf{f}^H_{\parallel} = -\frac{2\pi\mu}{\ln(L/r)}|(\mathbf{v}\cdot \mathbf{t})\mathbf{t}|,~~~~~\mathbf{f}^H_{\perp} = -\frac{4\pi\mu}{\ln(L/r)}|\mathbf{v}-(\mathbf{v}\cdot \mathbf{t})\mathbf{t}|
\end{equation}
where $\mathbf{f}^H_{\parallel}=(\mathbf{f}^H\cdot \mathbf{t})\mathbf{t}$ and $\mathbf{f}^H_{\perp}=\mathbf{f}^H-\mathbf{f}^H_{\parallel}$ are, respectively, tangential and orthogonal viscous drag components.
The coupling of liquid environment, filament mechanics and muscular activity provides a flexible platform to characterize biological locomotion at the microscopic scale (bacteria, protozoa, algae, etc.) and to design propulsion strategies in the context of artificial micro-swimmers \cite{Williams:2014,Park:2016}. 

\section{Applications}
\label{sec:applications}
We now proceed to illustrate the potential of our  framework with three different applications. We consider first a static problem in which self-contact, bending and twist give rise to the classic out-of-plane configurations denoted as plectonemes \cite{Ghatak:2005}, while the addition of stretching and shearing produces a different type of experimentally observed solutions, known as solenoids \cite{Ghatak:2005}. Then we turn our attention to two dynamic biophysical problems in which an active filament interacts with a solid and a liquid environment, exhibiting qualitatively different optimal biolocomotion strategies.

\subsubsection{Plectonemes and solenoids}
When an inextensible rod is clamped at one end and twisted a sufficiently large number of times at the other end, it becomes unstable, coils up and generates a characteristic structure known as {a} plectoneme \cite{Thompson:1996}. While this behavior has been well characterized both theoretically and experimentally \cite{Thompson:1996}, its analog for highly extensible filaments has been ignored. In particular, for large extensional and twisting strains qualitatively different  solutions arise, such as those corresponding to tightly packed solenoidal structures \cite{Ghatak:2005} whose properties are as yet poorly understood.

\begin{figure}
\begin{center}
\includegraphics[width=\textwidth]{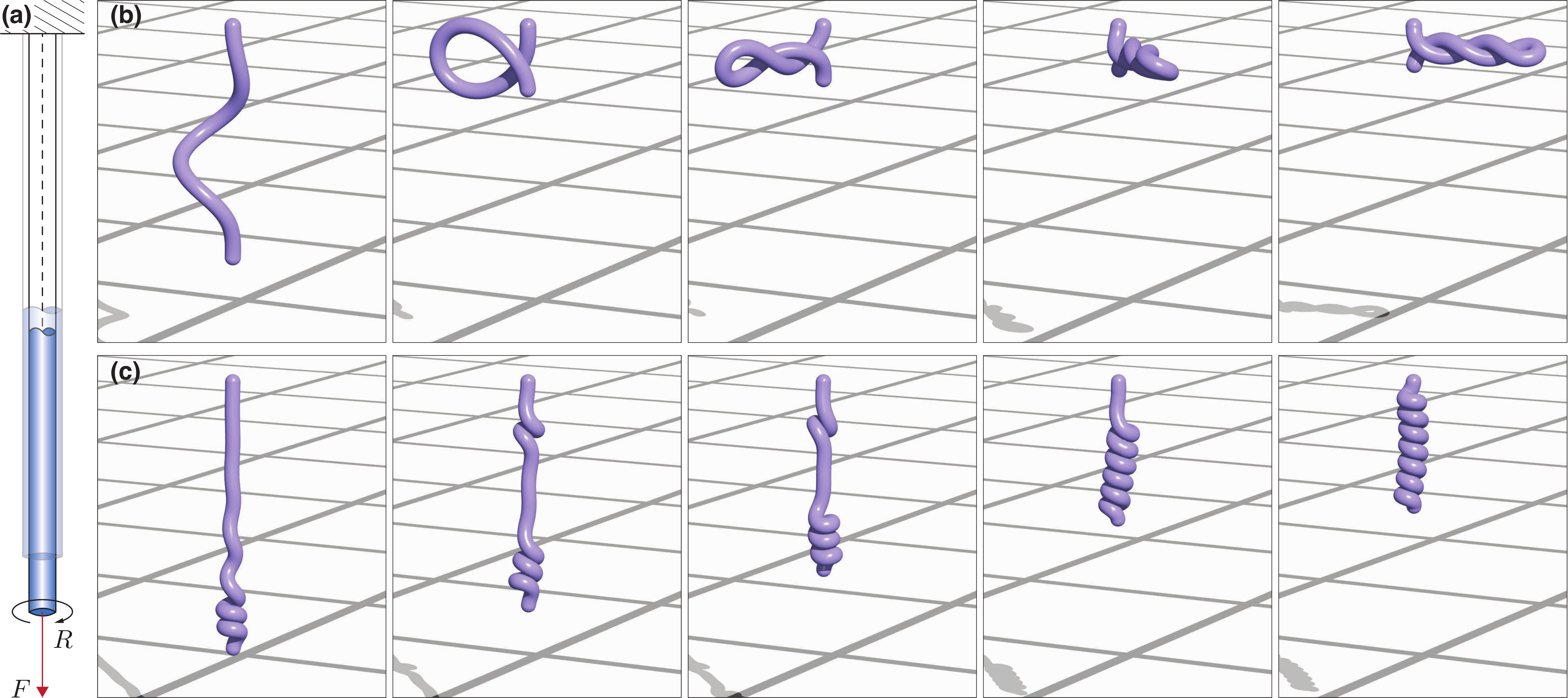}
\caption{\scriptsize{\textbf{Formation of plectonemes and solenoids.} (a) We consider a soft rod clamped at one end, subject to a constant vertical load $F$ and twisted $R$ times at the other end. (b) Formation of a plectoneme for $F=0$ (leading to the total elongation $L/\hat{L}\approx 1$) and $R=4$. (c) Formation of a solenoid for $F=300$~N (leading to the total elongation $L/\hat{L}\approx 1.15$) and $R=13$. Settings: length $L=1$~m, radius $r=0.025$~m, mass $m=1$~kg, Young's modulus $E=10^6$~Pa, shear modulus $G=2E/3$~Pa, shear/stretch matrix $\hat{\mathbf{S}}=\text{diag}(4G\hat{A}/3, 4G\hat{A}/3, E\hat{A})$~N, {bend}/twist matrix $\hat{\mathbf{B}}=\text{diag}(EI_1, EI_2, GI_3)$~Nm$^2$, dissipation constant $\gamma=2$~kg/(ms), $k_{sc}=10^4$~kg/s$^2$, $\gamma_{sc}=10$~kg/s, discretization elements $n=100$, timestep $\delta t = 0.01\delta l$~s, $T_{\text{twist}}=75$~s, $T_{\text{relax}}=50$~s.}}
\label{fig:solenoidsPlectonemes}
\end{center}
\end{figure}

Given the broad scope of our computational framework for the investigation of soft {filament} dynamics, we can now study the formation of both solenoids and plectonemes. As illustrated in Fig.~\ref{fig:solenoidsPlectonemes}a, a soft rod of Young's modulus $E=10^6$~Pa is clamped at one end, and subject to an axial load $F$, while also being twisted $R$ times at the other end. As experimentally and theoretically observed for $F=0$, i.e. in the absence of stretching ($L/\hat{L}\approx 1$), plectonemes are generated (Fig.~\ref{fig:solenoidsPlectonemes}b). When the load $F$ is increased so that the elongation of the rod approaches $L/\hat{L}\approx 1.15$, solenoids arise as predicted in \cite{Ghatak:2005} and illustrated in Fig.~\ref{fig:solenoidsPlectonemes}c. This test case, therefore, shows the ability of our solver to capture qualitatively different instability mechanisms, driven by the competition between the different modes of deformation of the rod. We leave the details of the explanation of the phase diagram for the formation of plectonemes, solenoids and intermediate structures \cite{Ghatak:2005} for a later study.

\subsubsection{Slithering}
\label{sec:limblessLocomotion}
The mechanics of slithering locomotion typical of snakes has been extensively investigated experimentally \cite{Gray:1950,Hu:2009,Transeth:2009}, theoretically \cite{Gray:1946,Guo:2008,Alben:2013} and computationally \cite{Erkmen:2002,Tanev:2005}. While biological experiments have provided quantitative insights, theoretical and computational models have been instrumental to characterize qualitatively the working principles underlying snake locomotion. Although these models implement different levels of realism, they generally rely on a number of key simplifications. Typically, theoretical models assume planar deformations \cite{Guo:2008} and/or disregard mechanics by prescribing body kinematics \cite{Alben:2013}. Computational models offer a more realistic representation, but they have mostly been developed for and tailored to robotic applications \cite{Erkmen:2002,Tanev:2005}. For example, snakes are often modeled as a relatively small set of hinges and/or springs representing pointwise localized actuators that connect contiguous rigid segments. Therefore, they do not account for the continuum nature of elastic body mechanics and biological muscular activity. Moreover, in robot replicas the critical feature of friction anisotropy is commonly achieved through the use of wheels \cite{Transeth:2009}. As a consequence computational models often assume only two sources of anisotropy, in the tangential and lateral direction with respect to the body. This is in contrast with biological experiments \cite{Hu:2009} that highlight the importance of all three sources of anisotropy, namely forward, backward and lateral.

\begin{figure}
\begin{center}
\includegraphics[width=\textwidth]{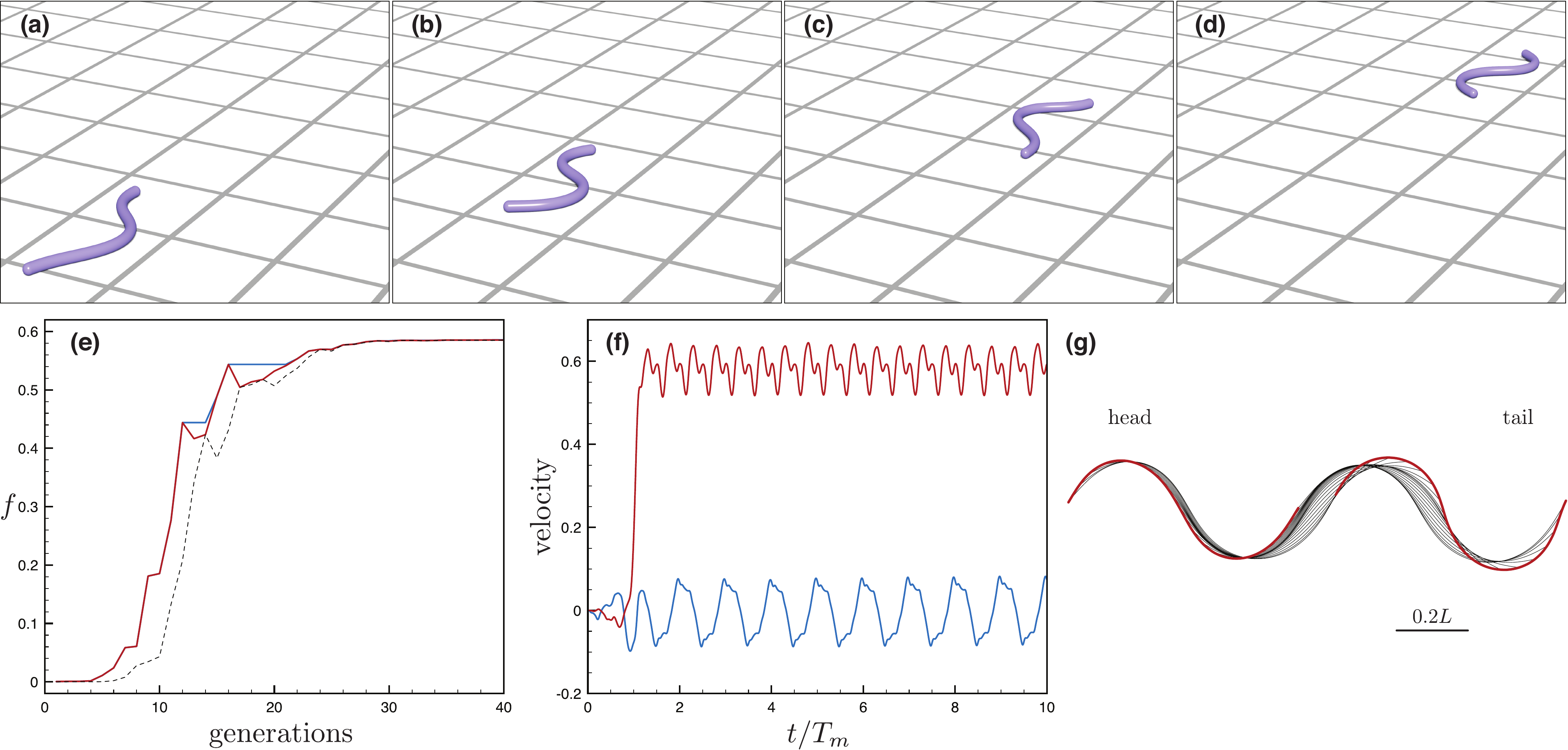}
\caption{\scriptsize{\textbf{Optimal lateral undulation gait.} (a, b, c, d) Instances at different times of a snake characterized by the identified optimal gait. (e) Evolution of the fitness function $f=v^{\text{fwd}}_{\text{max}}$ as function of the number of generations produced by CMA-ES. Solid blue, solid red and dashed black lines represent, respectively, the evolution of $f$ corresponding to the best solution, the best solution within the current generation, and the mean generation value. (f) Scaled forward (red) and lateral (blue) center of mass velocities versus normalized time. (g) Gait envelope over one oscillation period $T_m$. Red lines represent head and tail displacement in time. Settings: Froude number $Fr=0.1$, length $L=1$~m, radius $r=0.025$~m, density $\rho=10^3$~kg/m$^3$, $T_m=1$~s, Young's modulus $E=10^7$~Pa, shear modulus $G=2E/3$~Pa, shear/stretch matrix $\hat{\mathbf{S}}=\text{diag}(4G\hat{A}/3, 4G\hat{A}/3, E\hat{A})$~N, {bend}/twist matrix $\hat{\mathbf{B}}=\text{diag}(EI_1, EI_2, GI_3)$~Nm$^2$, dissipation constant $\gamma=5$~kg/(ms), gravity $g=9.81$~m/s$^2$, friction coefficient ratios $\mu^f_k:\mu^b_k:\mu^r_k=1:1.5:2$ and $\mu^f_s:\mu^b_s:\mu^r_s=1:1.5:2$ with $\mu^f_s=2\mu^f_k$, friction threshold velocity $v_{\epsilon}=10^{-8}$~m/s, ground stiffness and viscous dissipation $k_w=1$~kg/s$^2$ and $\gamma_w=10^{-6}$~kg/s, discretization elements $n=50$, timestep $\delta t = 2.5\cdot10^{-5}~T_m$, wavelength $\lambda_m= 0.97L$, phase shift $\phi_m=0$, torque B-spline coefficients $\beta_{i=0,\dots,5}=\{0,17.4, 48.5, 5.4, 14.7, 0\}$~Nm, bounds maximum attainable torque  $|\beta|^{\text{max}}_{i=0,\dots,5}=50$~Nm.}}
\label{fig:optimizedSnake}
\end{center}
\end{figure}

Our approach complements these previous attempts by accounting for physical and biological effects within a continuum framework (Eqs.~\ref{eq:velfinal}-\ref{eq:angmomentfinal}). In this section we demonstrate the qualitative and quantitative capabilities of the proposed method by reverse engineering optimal slithering gaits that maximize forward speed. 

We consider a soft filament of unit length actuated via a planar traveling torque wave of muscular activity in the direction perpendicular to the ground. The interaction with the substrate is characterized by the ratios $\mu^f_k:\mu^b_k:\mu^r_k=1:1.5:2$ and $\mu^f_s:\mu^b_s:\mu^r_s=1:1.5:2$ with $\mu^f_s=2\mu^f_k$, as experimentally observed for juvenile Pueblan milk snakes on a moderately rough surface \cite{Hu:2009}. The value of the friction coefficient $\mu_k^f$ is set so that the ratio between inertial and friction forces captured by the Froude number is $Fr=(L/T_m^2)/(\mu^f_k g)=0.1$, as measured for these snakes \cite{Hu:2009}. 

In the spirit of \cite{Gazzola:2012,Rees:2013,Rees:2015}, we wish to identify the fastest gaits by optimizing the filament muscular activity. The torque wave generated by the snake is parameterized according to Section~\ref{sec:muscularActivity} and is characterized by $N_m=6$ control points and a unit oscillation period $T_m$, so that overall we optimize for five parameters, four of which are responsible for the torque profile along the rod ($\beta_1$, $\beta_2$, $\beta_3$, $\beta_4$), while the last one represents the wavenumber $2\pi/\lambda_m$ (see Section~\ref{sec:muscularActivity}).

These parameters are {left} free to evolve from an initial zero value, guided by an automated optimization procedure that identifies the optimal values that maximize the snake's forward average speed $v^{\text{fwd}}_{\text{max}}$ over one activation cycle $T_m$. The algorithm of choice is the Covariance Matrix Adaptation - Evolution Strategy\cite{Hansen:2001,Hansen:2003} (CMA-ES) which has been proven effective in a range of biophysical and engineering problems, from the optimization of swimming gaits \cite{Gazzola:2012}, morphologies \cite{Rees:2013,Rees:2015} and collective dynamics \cite{Gazzola:2015a} to the identification of aircraft alleviation schemes \cite{Chatelain:2011} or virus traffic mechanisms \cite{Gazzola:2009}. The CMA-ES is a stochastic optimization algorithm that samples generations of $p$ parameter vectors from a multivariate Gaussian distribution $\mathcal{N}$. Here each parameter vector represents a muscular activation instance, and every generation entails the evaluation of $p=60$ different gaits. The covariance matrix of the distribution $\mathcal{N}$ is then adapted based on successful past gaits, chosen according to their corresponding cost function value $f=v^{\text{fwd}}_{\text{max}}$, until convergence to the optimum. 

The course of the optimization is reported in Fig.~\ref{fig:optimizedSnake} together with the kinematic details of the identified fastest gait. As can be noticed in Fig.~\ref{fig:optimizedSnake}e,f the forward scaled average speed approaches $v^{\text{fwd}}_{\text{max}}\simeq0.6$, consistent with experimental {evidence} \cite{Shine:2003}. Moreover, CMA-ES finds that the optimal wavelength is $\lambda_m\simeq L$ (Fig.~\ref{fig:optimizedSnake}g), again consistent with biological observations \cite{Hu:2009,Goldman:2010}. Thus, this value of wavelength strikes a balance between thrust production and drag minimization within the mechanical constraints of the system.

We note that a rigorous characterization of slithering locomotion would require the knowledge of a number of biologically relevant parameters (Young's and shear moduli of muscular tissue, maximum attainable torques, etc) and environmental conditions (terrain asperities, presence of pegs, etc) and goes beyond the scope of the present work. Nevertheless, this study illustrates the robustness, quantitative accuracy and suitability of our methodology for the characterization of bio-locomotion phenomena.

\subsubsection{Swimming}
We finally turn to apply the inverse design approach outlined in the previous section to the problem of swimming at low Reynolds numbers where viscous forces dominate inertial effects. We maintain the exact same set up as in the slithering case, while we change the environment from a solid substrate to a viscous fluid. The flow-filament interaction is then modeled via slender-body theory, as illustrated in Section~\ref{sec:hydro}.

Once again we inverse design planar optimal gaits for forward average speed $f=v^{\text{fwd}}_{\text{max}}$ within one  activation cycle $T_m$, by employing the same muscular activity parameterization as for slithering. In order to verify  \textit{a-posteriori} the biological relevance of the identified optimal solution, we consider the case of the sea urchin spermatozoon \textit{Echinus esculentus} \cite{Woolley:2001} which swims by means of helical or planar waves traveling along its flagella of length $L_s\simeq40$~$\mu$m. The gait corresponding to planar swimming is characterized by kinematic undulations of wavelength $\lambda_s<L_s$ and frequency $f_s\simeq2.8$. At $Re\simeq10^{-4}$ the spermatozoon attains the scaled velocity $v_s=U_s/(f_sL_s)\simeq0.08\pm0.03$, where $U_s$ is the dimensional cruise speed \cite{Woolley:2001}. Although this gait may not be the absolute optimal planar locomotion pattern, the fact that it is replicated in a large number of organisms \cite{Lauga:2009} suggests that it captures some effective features that we expect to qualitatively recover via our numerical optimization.

\begin{figure}
\begin{center}
\includegraphics[width=\textwidth]{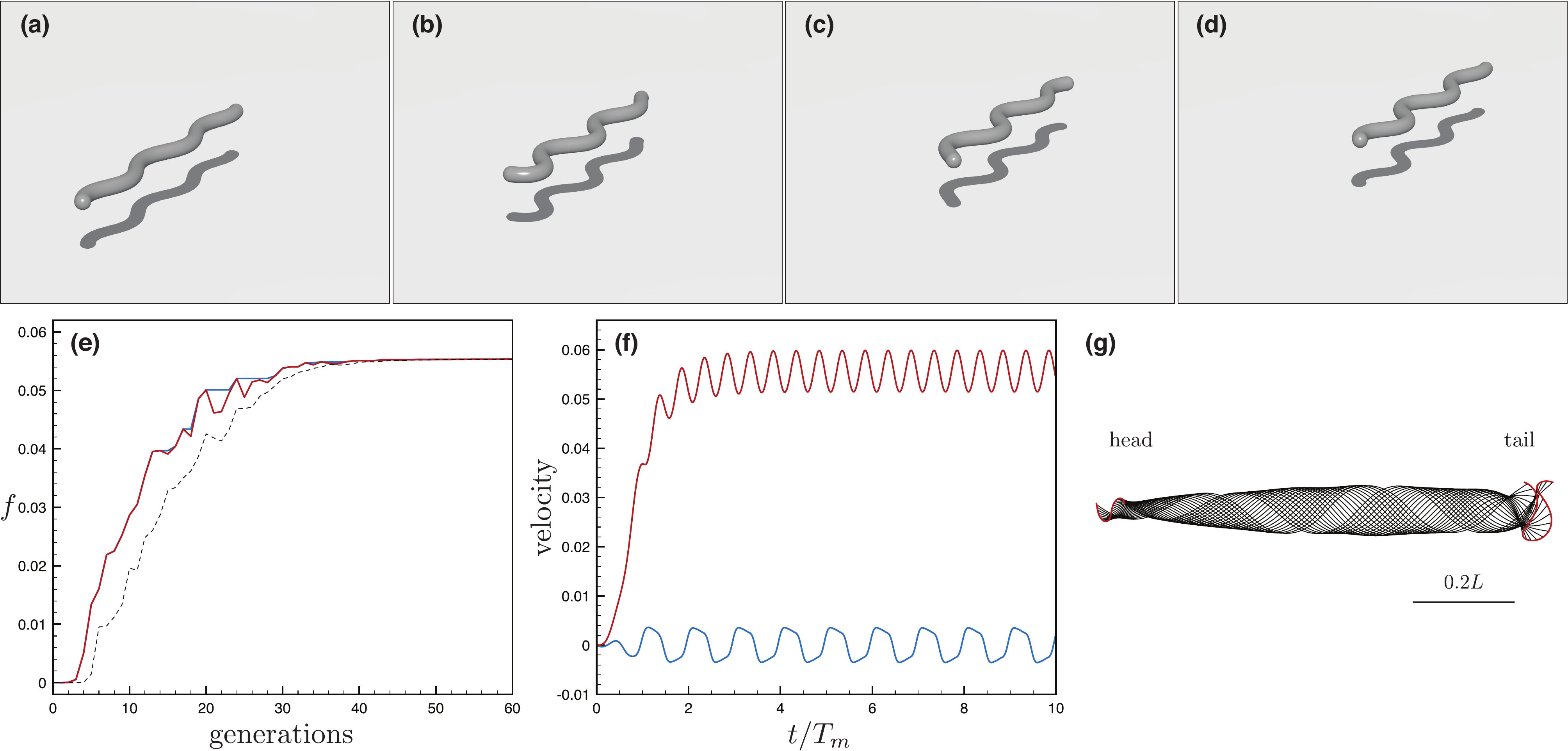}
\caption{\scriptsize{\textbf{Optimal planar swimming gait {at} low Reynolds number.} (a, b, c, d) Instances at different times of a filament swimming according to the identified optimal gait. (e) Evolution of the fitness function $f=v^{\text{fwd}}_{\text{max}}$ as function of the number of generations produced by CMA-ES. Solid blue, solid red and dashed black lines represent, respectively, the evolution of $f$ corresponding to the best solution, the best solution within the current generation, and the mean generation value. (f) Scaled forward (red) and lateral (blue) center of mass velocities versus normalized time. (g) Gait envelope over one oscillation period $T_m$. Red lines represent head and tail displacement in time. Settings: Reynolds number $Re=10^{-4}$, length $L=1$~m, radius $r=0.025$~m, filament density $\rho=10^3$~kg/m$^3$, $T_m=1$~s, Young's modulus $E=10^7$~Pa, shear modulus $G=2E/3$~Pa, shear/stretch matrix $\hat{\mathbf{S}}=\text{diag}(4G\hat{A}/3, 4G\hat{A}/3, E\hat{A})$~N, {bend}/twist matrix $\hat{\mathbf{B}}=\text{diag}(EI_1, EI_2, GI_3)$~Nm$^2$, dissipation constant $\gamma=5$~kg/(ms), discretization elements $n=50$, timestep $\delta t = 2.5\cdot10^{-5}~T_m$, wavelength $\lambda_m= 2.6L$, phase shift $\phi_m=0$, torque B-spline coefficients $\beta_{i=0,\dots,5}=\{0,50, 50, 50, 50, 0\}$~Nm, bounds maximum attainable torque  $|\beta|^{\text{max}}_{i=0,\dots,5}=50$~Nm.}}
\label{fig:optimizedStokesLateral}
\end{center}
\end{figure}

The course of the optimization is reported in Fig.~\ref{fig:optimizedStokesLateral} together with the kinematic details of the identified fastest gait. As can be noticed in Fig.~\ref{fig:optimizedStokesLateral}e,f the forward average scaled speed and wavelength approach $v^{\text{fwd}}_{\text{max}}\simeq0.055$ and  $\lambda_m\simeq 0.38L$, qualitatively and quantitatively consistent with experimental {evidence} \cite{Woolley:2001}.

As in the previous section, we note that a rigorous characterization of swimming at low Reynolds numbers would require the knowledge of a number of biologically relevant parameters and environmental conditions, and goes beyond the scope of the present work. Nevertheless, this and the previous study illustrate how the interplay between filament mechanics and the surrounding environment crucially affects propulsive gaits, as is biologically evident and automatically recovered via our numerical inverse design approach.

\section{Conclusions}
We have presented a {robust and flexible framework} for the simulation of soft filaments deforming in three dimensional space. Our scheme accounts at any given cross section for all possible deformation degrees of freedom, namely normal and orthonormal {bending, twisting, stretching and shearing}. Furthermore, we enhance it to handle self-contact, muscular activity, solid boundaries, isotropic and anisotropic friction as well as hydrodynamics. The outcome is a relatively simple algorithm able to simulate a plethora of physical and biological phenomena.

We validate the proposed method {against} a battery of benchmark problems entailing different physical aspects and boundary conditions, and we examine its convergence properties in depth. We further showcase the capabilities of our approach by studying several applications: the formation of solenoids and plectonemes and the evolutionary optimization of terrestrial limbless locomotion and swimming. {We emphasize that {using} an evolutionary strategy in combination with a numerical solver severely {tests} the robustness of the solver itself, due to the variety of candidate solutions produced throughout the process.}

{Therefore, our} results demonstrate the robustness and accuracy of our discrete model under a variety of different deformation regimes and illustrate its flexibility and potential for a wide range of passive and active applications involving soft filaments.

Ongoing work involves its coupling to realistic high Reynolds number flow solvers \cite{Gazzola:2011a} as well as its integration with sensory feedback models for the characterization of locomotory neural circuitry \cite{Gazzola:2015}.

\section*{Acknowledgements}
We thank Nicholas Charles for a careful reading of and comments on the paper. We thank the Blue Waters project (OCI-0725070, ACI-1238993), a joint effort of the University of Illinois at Urbana-Champaign and its National Center for Supercomputing Applications, for partial support (MG).

\appendix

\section{Lagrangian governing equations}

The following shows the conversion of governing equations for {\it unstretchable} filaments from the lab to material frame of reference. Note that in the Governing Equations and Numerical Methods sections of the main text all terms are scaled appropriately by the local dilatation to account for stretching.

The time and space derivatives of the centerline $\mathbf{r}(s,t)$ are associated with the velocity $\mathbf{v}$ and tangent field $\mathbf{t}$
\begin{equation}
\mathbf{v}=\frac{\partial\mathbf{r}}{\partial t}, ~~~~~\mathbf{t}=\frac{\partial\mathbf{r}}{\partial s},
\label{eq:velocities}
\end{equation}
with $|\mathbf{t}|=1$, since $s$ is the current arc-length.

Similarly, time and space derivatives of the material frame $\mathbf{Q}$, due to the orthonormality of the directors, are associated by definition with the angular velocity $\boldsymbol{\omega}$ and generalized curvature $\boldsymbol{\kappa}$ vectors, so that
\begin{eqnarray}
\frac{\partial\mathbf{d}_j}{\partial t} &=& \frac{\partial(\mathbf{Q}^T\mathbf{e}_j)}{\partial t} = \frac{\partial\mathbf{Q}^T}{\partial t}\mathbf{e}_j = \frac{\partial\mathbf{Q}^T}{\partial t}\mathbf{Q}\mathbf{d}_j = \boldsymbol{\omega}\times\mathbf{d}_j, ~~~~~j=1,2,3
\label{eq:transport1}
\end{eqnarray}
\begin{eqnarray}
\frac{\partial\mathbf{d}_j}{\partial s} &=& \frac{\partial(\mathbf{Q}^T\mathbf{e}_j)}{\partial s} = \frac{\partial\mathbf{Q}^T}{\partial t}\mathbf{e}_j = \frac{\partial\mathbf{Q}^T}{\partial s}\mathbf{Q}\mathbf{d}_j = \boldsymbol{\kappa}\times\mathbf{d}_j, ~~~~~j=1,2,3
\label{eq:transport2}
\end{eqnarray}
where the equivalences $\partial_t\mathbf{Q}^T\cdot\mathbf{Q} = \boldsymbol{\omega} \times (\cdot)$ and $\partial_s\mathbf{Q}^T\cdot\mathbf{Q} = \boldsymbol{\kappa} \times (\cdot)$ hold. These kinematic equations combined with the linear and angular momentum balance laws at a cross section \cite{Landau:1959b} yield the governing equations for the Cosserat rod 
\begin{eqnarray}
\frac{\partial \mathbf{r}}{\partial t} &=& \mathbf{v}\label{eq:vel} \\
\frac{\partial \mathbf{d}_j}{\partial t} &=& \boldsymbol{\omega} \times \mathbf{d}_j,~~~~~j=1,2,3\label{eq:frame}\\
\frac{\partial (\rho A \mathbf{v})}{\partial t} &=& \frac{\partial \mathbf{n}}{\partial s}+ \mathbf{f}\label{eq:linmoment}\\
\frac{\partial \mathbf{h}}{\partial t} &=& \frac{\partial \boldsymbol{\tau}}{\partial s}+ \frac{\partial \mathbf{r}}{\partial s}\times \mathbf{n}+\mathbf{c}\label{eq:angmoment},
\end{eqnarray}
where $\rho$ is the constant material density, $A$ is the cross sectional area in its current state (so that $\rho A \mathbf{v}$ is the linear momentum line density), $\mathbf{h}(s,t)$ is the angular momentum line density, $\mathbf{n}(s,t)$ and $\boldsymbol{\tau}(s,t)$ are the internal force and torque resultants, and $\mathbf{f}$ and $\mathbf{c}$ are external body force and torque line densities.

The internal force $\mathbf{n}$ and torque $\boldsymbol{\tau}$ resultants depend on the geometric and material properties of the filament. This dependence is embedded via the material or constitutive laws, which must be frame invariant, rendering their definition in a Lagrangian setting most natural. Moreover, we note that the angular momentum line density can be readily expressed in the material frame as $\mathbf{Q}\mathbf{h}=\mathbf{h}_{\mathcal{L}}=\rho \mathbf{I}\boldsymbol{\omega}_{\mathcal{L}}$, where the tensor $\mathbf{I}$ is the second area moment of inertia which, assuming circular cross sections, takes the form
\[
\mathbf{I} = \left( \begin{array}{ccc}
I_1 & 0 & 0 \\
0 & I_2 & 0 \\
0 & 0 & I_3 \end{array} \right)=\frac{A^2}{4\pi}\left( \begin{array}{ccc}
1 & 0 & 0 \\
0 & 1 & 0 \\
0 & 0 & 2 \end{array} \right)=\frac{A^2}{4\pi}\text{diag(1,1,2)}.
\]

To derive the Lagrangian form of the angular momentum equation, we begin with the definition of the material frame

\begin{eqnarray}
\mathbf{d}_j = \mathbf{Q}^T\mathbf{e}_j, ~~~~~j=1,2,3
\end{eqnarray}

and definitions of time and space derivatives of its frame directors
\begin{eqnarray}
\frac{\partial\mathbf{d}_j}{\partial t} &=& \boldsymbol{\omega}\times\mathbf{d}_j, ~~~~~j=1,2,3 \\
\frac{\partial\mathbf{d}_j}{\partial s} &=& \boldsymbol{\kappa}\times\mathbf{d}_j, ~~~~~j=1,2,3.
\end{eqnarray}

We can use these definitions to establish relationships between time and space derivatives of the laboratory and material frames.
\begin{align}
\frac{\partial \mathbf{x}}{\partial t} &= \frac{\partial \big( \mathbf{Q}^T \mathbf{x}_{\mathcal{L}}\big)}{\partial t} \\
&= \frac{\partial \big( \sum_{j=1}^3 \mathbf{d}_j \mathbf{x}_{\mathcal{L}j}\big)}{\partial t} \\
&= \sum_{j=1}^3 \partial_t \big( \mathbf{d}_j \mathbf{x}_{\mathcal{L}j} \big) \\
&= \sum_{j=1}^3 \big( \partial_t \mathbf{d}_j \big)\mathbf{x}_{\mathcal{L}j} + \sum_{j=1}^3 \mathbf{d}_j \big( \partial_t \mathbf{x}_{\mathcal{L}j} \big) \\
&= \sum_{j=1}^3 \big( \boldsymbol{\omega} \times \mathbf{d}_j \big) \mathbf{x}_{\mathcal{L}j} + \sum_{j=1}^3 \mathbf{d}_j \big( \partial_t \mathbf{x}_{\mathcal{L}j} \big) \\
\frac{\partial \mathbf{x}}{\partial t} &= \boldsymbol{\omega} \times \big( \mathbf{Q}^T \mathbf{x}_{\mathcal{L}}\big) + \mathbf{Q}^T \frac{\mathbf{x}_{\mathcal{L}}}{\partial t} \label{eq:time_deriv_app} \\
\frac{\partial \mathbf{x}}{\partial s} &= \boldsymbol{\kappa} \times \big( \mathbf{Q}^T \mathbf{x}_{\mathcal{L}}\big) + \mathbf{Q}^T \frac{\mathbf{x}_{\mathcal{L}}}{\partial s} \label{eq:space_deriv_app}
\end{align}
Because the second moment of inertia is defined in the material frame, the governing equation for the angular momentum is most naturally expressed in the material frame as well ($\mathbf{I}$ in the lab frame is a function of space and time rendering its use cumbersome, while $\mathbf{I}_{\mathcal{L}}$ is a constant and, in our isotropic case, diagonal matrix in the material frame). We can use Eqs.~\eqref{eq:time_deriv_app} and \eqref{eq:space_deriv_app} to convert all of the terms in Eq.~\eqref{eq:angmoment} to the material frame.
\begin{eqnarray}
\frac{\partial (\rho\mathbf{I}\boldsymbol{\omega})}{\partial t} = \mathbf{Q}^T \frac{\partial (\rho\mathbf{I}\boldsymbol{\omega})_{\mathcal{L}}}{\partial t} + \boldsymbol{\omega} \times \big( \mathbf{Q}^T (\rho\mathbf{I}\boldsymbol{\omega})_{\mathcal{L}} \big) \quad \text{LHS} \\
\frac{\partial \boldsymbol{\tau}}{\partial s} = \mathbf{Q}^T \frac{\partial \boldsymbol{\tau}_{\mathcal{L}}}{\partial t} + \boldsymbol{\omega} \times \left( \mathbf{Q}^T \boldsymbol{\tau}_{\mathcal{L}} \right) \quad \text{RHS} \\
\frac{\partial \mathbf{r}}{\partial s}\times \mathbf{n} = \mathbf{Q}^T \left( \frac{\partial \mathbf{r}}{\partial s}\right)_{\mathcal{L}} \times \mathbf{Q}^T \mathbf{n}_{\mathcal{L}} \quad \text{RHS}\\
\mathbf{c} = \mathbf{Q}^T \mathbf{c}_{\mathcal{L}} \quad \text{RHS}
\end{eqnarray}
Multiplying by $\mathbf{Q}$ and noting that $\mathbf{I}\boldsymbol{\omega} = \mathbf{Q}^T \mathbf{I}_{\mathcal{L}}\boldsymbol{\omega}_{\mathcal{L}}$ yields our final governing equation for the angular momentum, expressed in the material frame
\begin{eqnarray}
\frac{\partial (\rho\mathbf{I}_{\mathcal{L}}\boldsymbol{\omega}_{\mathcal{L}})}{\partial t} &=& \frac{\partial \boldsymbol{\tau}_{\mathcal{L}}}{\partial s} + \boldsymbol{\kappa}_{\mathcal{L}}\times\boldsymbol{\tau}_{\mathcal{L}}+ \mathbf{Q}\frac{\partial \mathbf{r}}{\partial s}\times \mathbf{n}_{\mathcal{L}}+(\rho\mathbf{I}_{\mathcal{L}}\boldsymbol{\omega}_{\mathcal{L}})\times\boldsymbol{\omega}_{\mathcal{L}} + \mathbf{c}_{\mathcal{L}}.
\end{eqnarray}
Aside from the balance of angular momentum, converting Eqs.~\eqref{eq:frame} and \eqref{eq:linmoment} to the material frame requires simply a direct application of the definition of the material frame to the RHS quantities.

\section{Time discretization}
In order to advance the discrete Hamiltonian system of Eqs.~(\ref{eq:velfinaldiscr}-\ref{eq:angmomentfinaldiscr}), we choose the energy conserving, second order position Verlet time integration scheme. This allows us to write a full iteration from time $t$ to $t+\delta t$ as
\begin{eqnarray}
\mathbf{r}_i\left(t+\frac{\delta t}{2}\right)&=& \mathbf{r}_i(t)+\frac{\delta t}{2}\cdot\mathbf{v}_i(t), \hspace{5cm}i=[1,n+1]\label{eq:xupdate1}\\
\mathbf{Q}_i\left(t+\frac{\delta t}{2}\right)&=& \exp{\left[\frac{\delta t}{2}\boldsymbol{\omega}^i_{\mathcal{L}}(t)\right]}\cdot \mathbf{Q}_i(t),\hspace{4.7cm}i=[1,n]\label{eq:qupdate1}\\
\mathbf{v}_i(t+\delta t)&=& \mathbf{v}_i(t)+\delta t\cdot\frac{d\mathbf{v}_i}{d t}\left(t+\frac{\delta t}{2}\right), \hspace{3.7cm}i=[1,n+1]\\
\boldsymbol{\omega}^i_{\mathcal{L}}(t+\delta t)&=& \boldsymbol{\omega}^i_{\mathcal{L}}(t)+\delta t \cdot \frac{d \boldsymbol{\omega}^i_{\mathcal{L}}}{d t}\left(t+\frac{\delta t}{2}\right), \hspace{4cm}i=[1,n]\label{eq:accupdate}\\
\mathbf{r}_i(t+\delta t)&=& \mathbf{r}_i\left(t+\frac{\delta t}{2}\right)+\frac{\delta t}{2}\cdot\mathbf{v}_i\left(t+\frac{\delta t}{2}\right)\label{eq:xupdate2}\hspace{3cm}i=[1,n+1]\label{eq:angaccupdate}\\
\mathbf{Q}_i(t+\delta t)&=& \exp{\left[\frac{\delta t}{2}\boldsymbol{\omega}^i_{\mathcal{L}}\left(t+\frac{\delta t}{2}\right)\right]}\cdot \mathbf{Q}_i\left(t+\frac{\delta t}{2}\right).\hspace{2.7cm}i=[1,n]\label{eq:qupdate2}
\end{eqnarray}
The time integrator is structured in three blocks. A first half step position update (Eqs.~\ref{eq:xupdate1}, \ref{eq:qupdate1}), followed by the evaluation of local accelerations (Eqs.~\ref{eq:accupdate}, \ref{eq:angaccupdate}), and finally the second half step position update (Eqs.~\ref{eq:xupdate2}, \ref{eq:qupdate2}). Therefore, the second order position Verlet scheme entails only one right hand side evaluation of Eqs.~(\ref{eq:linmomentfinaldiscr}, \ref{eq:angmomentfinaldiscr}), the most computationally expensive operation. We also emphasize that for the numerical integration of Eqs.~(\ref{eq:qupdate1}, \ref{eq:qupdate2}) the Rodrigues formula is employed, implying the direct use of $\boldsymbol{\omega}_{\mathcal{L}}$ instead of $\boldsymbol{\omega} = \mathbf{Q}^T\boldsymbol{\omega}_{\mathcal{L}}$. By foregoing an implicit integration scheme we can quickly incorporate a number of additional physical effects and soft constraints. This algorithm strikes a balance between computing costs, numerical accuracy and implementation modularity.

\section{Discrete operators}

The discrete operators $\Delta^h:\{\mathbb{R}^3\}_N\rightarrow\{\mathbb{R}^3\}_{N+1}$ and $\mathcal{A}^h:\{\mathbb{R}^3\}_N\rightarrow\{\mathbb{R}^3\}_{N+1}$ take the form
\[
\mathbf{y}_{j=1,\dots,N+1}=
\Delta^h(\mathbf{x}_{i=1,\dots,N})=
\begin{dcases}
    \mathbf{x}_1& \text{if } j=1\\
    \mathbf{x}_j - \mathbf{x}_{j-1}& \text{if } 1<j\le N\\
    -\mathbf{x}_N& \text{if } j=N+1\\
\end{dcases},
\]
\[
\mathbf{y}_{j=1,\dots,N+1}=
\mathcal{A}^h(\mathbf{x}_{i=1,\dots,N})=
\begin{dcases}
    \frac{\mathbf{x}_1}{2}& \text{if } j=1\\
    \frac{\mathbf{x}_j + \mathbf{x}_{j-1}}{2}& \text{if } 1<j\le N\\
    \frac{\mathbf{x}_N}{2}& \text{if } j=N+1\\
\end{dcases}.
\]
We note that $\Delta^h$ and $\mathcal{A}^h$ operate on a set of $N$ vectors and return $N+1$ vectors of differences or convert integrated quantities to point-wise, respectively, consistent with Eqs.~(\ref{eq:linmomentfinaldiscr}, \ref{eq:angmomentfinaldiscr}).


\section{Derivation of the vertical displacement of a Timoshenko cantilever beam}
\begin{figure}
\begin{center}
\includegraphics[width=0.75\textwidth]{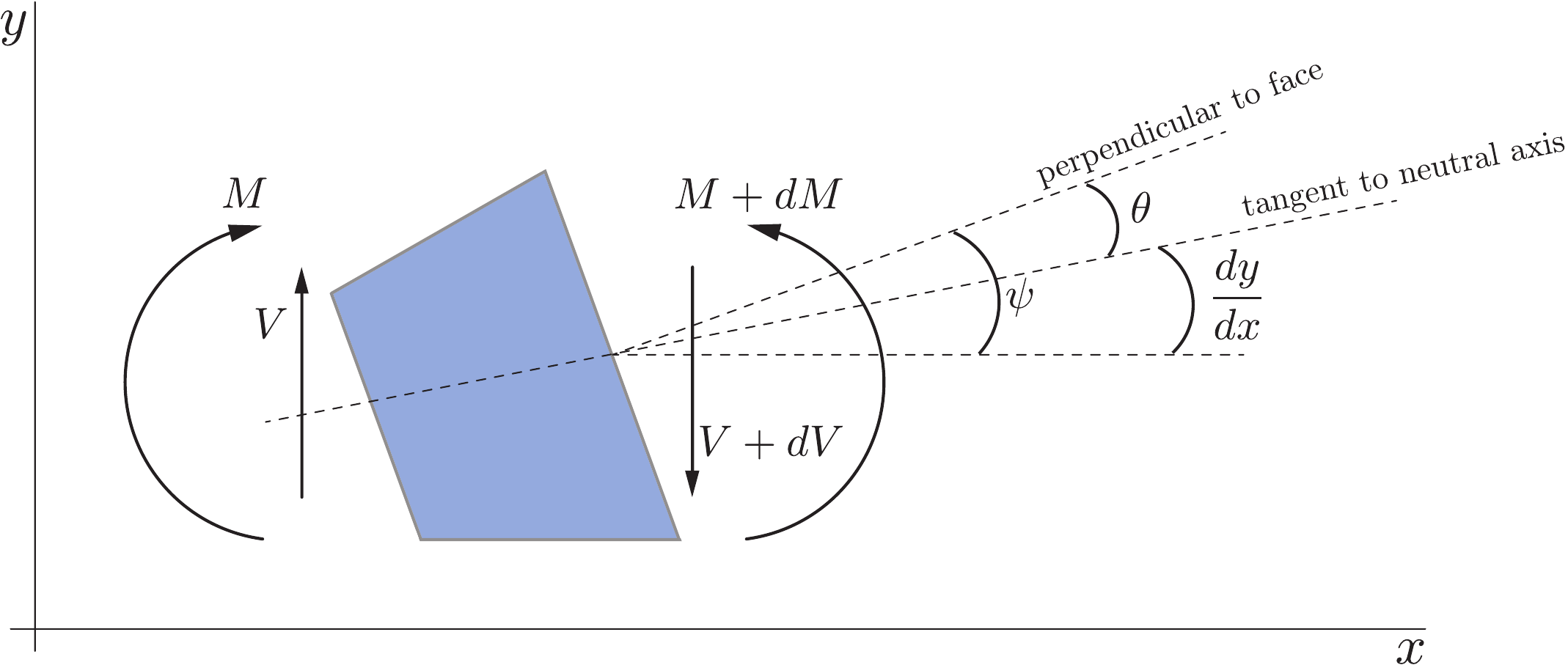}
\caption{\scriptsize{\textbf{Free body diagram of an infinitesimal beam element.} The sketch represents an infinitesimal element undergoing bending and shear deformations, assuming small deflections so that $x \simeq s$. The bending moment is denoted as $M$, the shear force as $V$, the vertical displacement as $y$. The bending angle $\psi$, the shear angle $\theta$ and the slope of the vertical displacement $d_x y$ are related to each other so that $\psi=\theta+d_x y$.}}
\label{fig:freeBodyDiagram_app}
\end{center}
\end{figure}

We briefly derive here the analytical expression of the vertical displacement $y$ of Eq.~(\ref{eq:displacementTimo}) for the cantilever beam problem of Fig.~5a.
In order to do so we make use of the free body diagram of Fig.~\ref{fig:freeBodyDiagram_app}, and of the constitutive relations of Table~\ref{tab:strains}. Moreover, we disregard axial extension and assume small deformations so that the coordinate $x$ (along the direction $\mathbf{k}$) is approximated by the arc-length $s$.

Recalling that the bending strain is the space derivative of the bending angle $\psi$ (Fig.~\ref{fig:freeBodyDiagram_app}), we may write
\begin{eqnarray}
\frac{\partial \psi}{\partial s}=\frac{M}{EI_1},\label{eq:beamElastic_app}
\end{eqnarray}
where $M$ is the bending moment, $E$ the Young's modulus, and $I_1$ is the second area moment of inertia about the axis $\mathbf{j}=\mathbf{k}\times\mathbf{i}$ (see Fig.~5a). The shear angle $\theta$, as illustrated in Fig.~\ref{fig:freeBodyDiagram_app}, is the difference between the bending angle and the slope of the centerline, so that
\begin{eqnarray}
\psi - \frac{\partial y}{\partial s}=\frac{V}{\alpha_cAG}\label{eq:beamShear_app},
\end{eqnarray}
where $V$ is the shear force, $A$ is the cross sectional area, $G$ is the shear modulus, and $\alpha_c$ is the Timoshenko factor \cite{Gere:2001}. If a point load $F$ is applied downward at $s=L$, where $L$ is the length of the rod, a free body diagram of the beam yields $M = -F(L-s)$ and $V=F$, so that 
\begin{eqnarray}
\frac{\partial \psi}{\partial s}&=&-\frac{F(L-s)}{EI_1},\label{eq:beamElastic2_app}\\
\psi - \frac{\partial y}{\partial s}&=&\frac{F}{\alpha_cAG}\label{eq:beamShear2_app}
\end{eqnarray}

By integrating Eq.~(\ref{eq:beamElastic2_app}) with boundary conditions $\psi=0$ at $s=0$, injecting the solution $\psi$ into Eq.~(\ref{eq:beamShear2_app}), and integrating again with boundary conditions $y=0$ at $s=0$, we obtain 
\begin{equation}
y=-\frac{F}{\alpha_c\hat{A}G}\hat{s} - \frac{F\hat{L}}{2E\hat{I}_1}\hat{s}^2 + \frac{F}{6E\hat{I}_1}\hat{s}^3.
\end{equation}

\section{Further Validations}

\subsection{Euler buckling instability}
\begin{figure}
\begin{center}
\includegraphics[width=\textwidth]{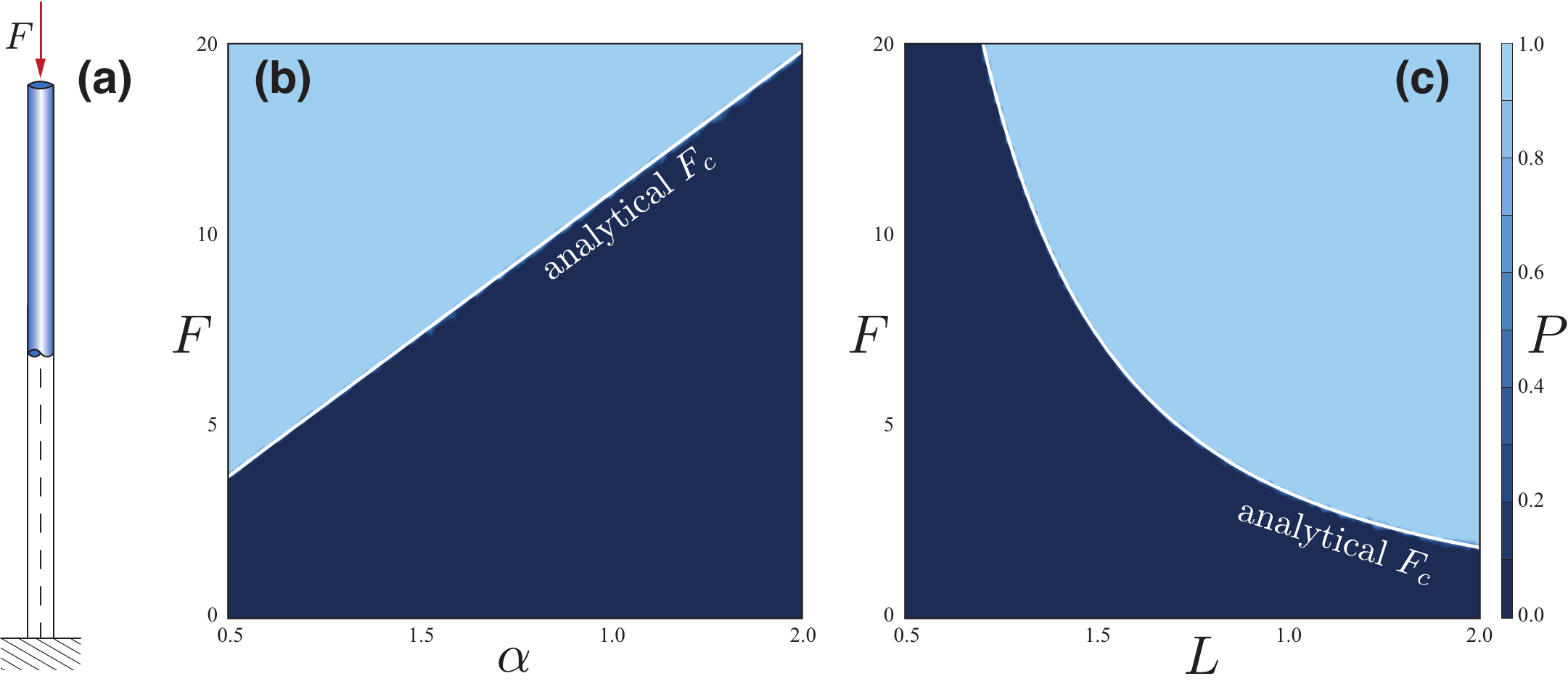}
\caption{\scriptsize{\textbf{Euler buckling instability.} (a) Inextensible and unshearable rod vertically initialized and subject to the axial load $F$. (b) Probability $P$ of observing an instability event as a function of the compression force $F$ and the bending rigidity $\alpha$ for a fixed length $L=1$~m. The corresponding analytical solution is overlaid as reference (white line). (c) Probability $P$ of observing an instability event as a function of the compression force $F$ and the length $L$ for a fixed bending rigidity $\alpha=1$~Nm$^2$. The corresponding analytical solution is overlaid as reference (white line). The probability $P$ is determined by performing ten simulations for each pair of parameters $(F, \alpha)$ or $(F, L)$. Each simulation is initially perturbed by applying to every discretization node a small random force sampled from a uniform distribution, such that $\|F_R\|\sim\mathcal{U}(0,10^{-2})$~N. The occurrence of an instability is detected whenever the rod bending energy $E_B>E_{th}$ with $E_{th}=10^{-3}$ J. Settings: rod's mass $m_r=1$~kg, twist stiffness $\beta=2\alpha/3$~Nm$^2$, shear/stretch matrix $\mathbf{S}=10^5\cdot\pmb{1}$~N, bend/twist matrix $\mathbf{B}=\text{diag}(\alpha, \alpha, \beta)$~Nm$^2$, radius $r=0.025L$~m, discretization elements $n=100$~m$^{-1}$, timestep $\delta t = 10^{-5}$~s, simulation time $T_{\text{sim}}=10$~s, dissipation constant $\gamma=0$.}}
\label{fig:eulerBuckling}
\end{center}
\end{figure}

Euler buckling involves a single straight isotropic, inextensible and unshearable rod subject to an axial load $F$, as depicted in Fig.~\ref{fig:eulerBuckling}a. The critical axial load $F_c$ that the rod can withstand before bending can be expressed analytically \cite{Love:1906} as 
\begin{equation}
F_c=\frac{\pi^2 E I}{(bL)^2},
\end{equation}
where $E$ is the modulus of elasticity of the rod, $I=I_1=I_2$ is the area moment of inertia, $L$ is the length, and $b$ is a constant which depends on the boundary conditions. If both ends are fixed in space but free to rotate then $b=1$, thus
\begin{equation}
F_c=\pi^2\frac{\alpha}{L^2},
\label{eq:criticalEulerForce}
\end{equation}
where $\alpha=EI$ is the bending stiffness.

We test our solver against the above analytical solution by simulating the time evolution of an inextensible and unshearable rod. In Fig.~\ref{fig:eulerBuckling}a, we show the results of our computations in the limit when both ends are free to rotate and all their spatial degrees of freedom are fixed except for one which allows the top end to slide vertically. The inextensible and unshearable conditions are enforced numerically by setting the entries of the matrix $\mathbf{S}$ to be much larger than those of $\mathbf{B}$ (details in Fig.~\ref{fig:eulerBuckling}).

We explore the phase spaces $F$-$\alpha$ and $F$-$L$ and determine the probability of a randomly perturbed rod to undergo a buckling event, characterized by the bending energy exceeding the small threshold value $E_B>E_{th}$. As can be seen in Fig.~\ref{fig:eulerBuckling}b,c the obtained probability landscapes exhibit a sharp transition in agreement with the exact solution of Eq.~(\ref{eq:criticalEulerForce}).

\subsection{Mitchell buckling instability}

The Euler buckling benchmark allows us to test the capability of our solver to capture the onset of an instability relative to the bending modes. Next  we consider the Mitchell buckling instability that simultaneously accounts for both bend and twist. When the ends of an isotropic, inextensible and unshearable filament are joined together, the resulting configuration is a planar circular shape. If the two ends are twisted by a given angle and glued together, a circular shape with uniform twist density is obtained. When the total twist $\Phi$, i.e. the integrated twist line density along the filament, exceeds a critical value $\Phi_c$ the rod buckles out of plane. This critical total twist can be expressed analytically \cite{Goriely:2006} as 
\begin{equation}
\Phi_c=\frac{2\sqrt{3}\pi} {\beta/\alpha},
\label{eq:criticalMitchell}
\end{equation}
where $\alpha$ and $\beta$ are, respectively, the bending and twist rigidities.

\begin{figure}
\begin{center}
\includegraphics[width=\textwidth]{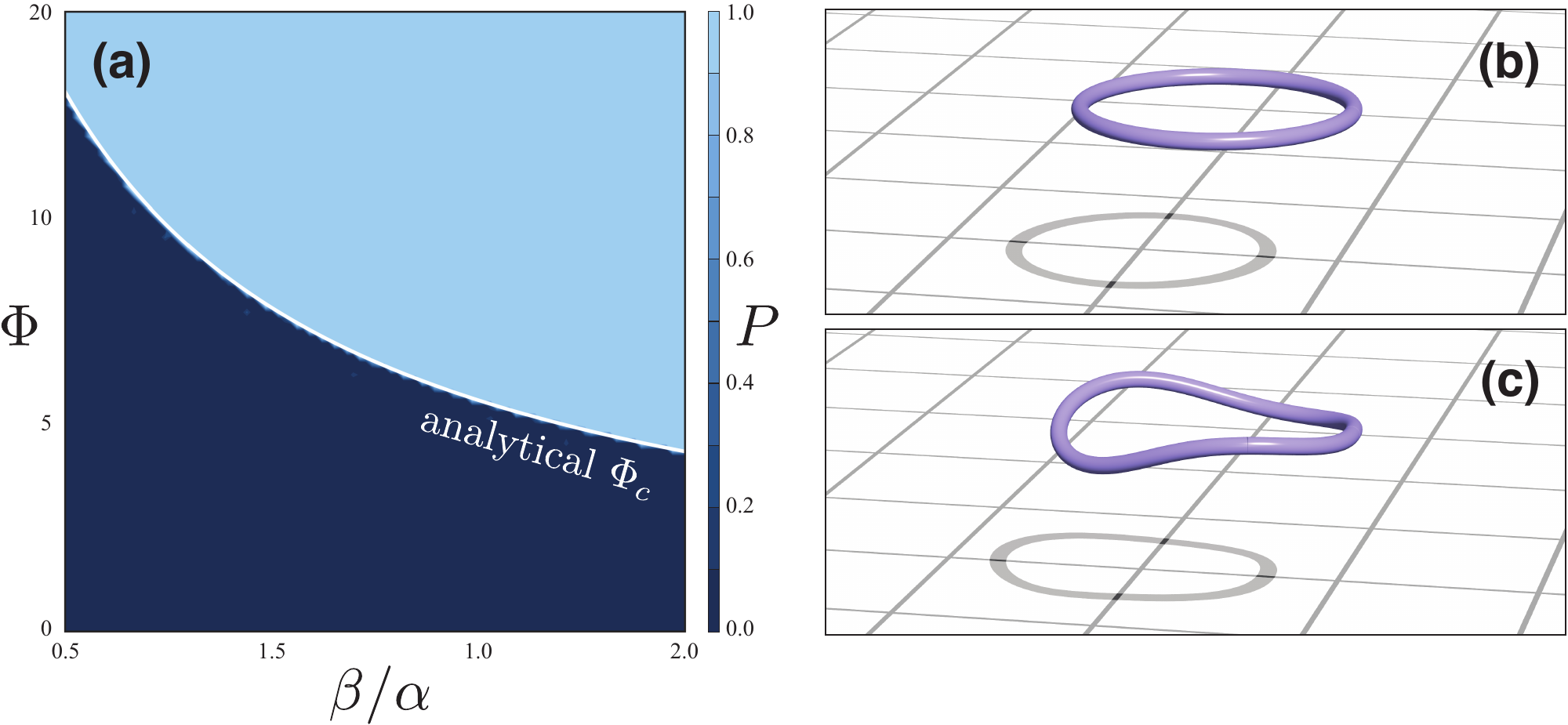}
\caption{\scriptsize{\textbf{Mitchell buckling instability.} (a) Probability $P$ of observing an instability event as a function of the total twist $\Phi$ and the ratio between bending and twist rigidities $\beta/\alpha$. The corresponding analytical solution is overlaid as reference (white line). The probability $P$ is determined by performing ten simulations for each pair of parameters $(\Phi, \beta/\alpha)$. Each simulation is initially perturbed by applying to every discretization node a small random force sampled from a uniform distribution, such that $\|F_R\|\sim\mathcal{U}(0,10^{-3})$~N. The occurrence of an instability is detected whenever the rod translational energy $E_T>E_{th}$ with $E_{th}=10^{-4}$ J. (b-c) Visualization of a Mitchell bucking instability event for $\Phi=10$ and $\beta/\alpha=2$. Settings: rod's mass $m_r=1$~kg, length $L=1$~m, bending stiffness $\alpha=1$~Nm$^2$, shear/stretch matrix $\mathbf{S}=10^5\cdot\pmb{1}$~N, bend/twist matrix $\mathbf{B}=\text{diag}(\alpha, \alpha, \beta)$~Nm$^2$, radius $r=0.01L$~m, discretization elements $n=50$~m$^{-1}$, timestep $\delta t = 10^{-5}$~s, simulation time $T_{\text{sim}}=2$~s, dissipation constant $\gamma=0$.}}
\label{fig:mitchellBuckling}
\end{center}
\end{figure}

We explore the phase space $F$-$(\beta/\alpha)$ and determine the probability of a randomly perturbed rod to undergo a buckling event, characterized by the translational energy exceeding a small threshold value $E_T>E_{th}$. As can be seen in Fig.~\ref{fig:mitchellBuckling}, our results show a sharp transition in agreement with the exact solution of Eq.~(\ref{eq:criticalMitchell}).

\subsection{Twist forced vibrations in a stretched rod}
\begin{figure}
\begin{center}
\includegraphics[width=\textwidth]{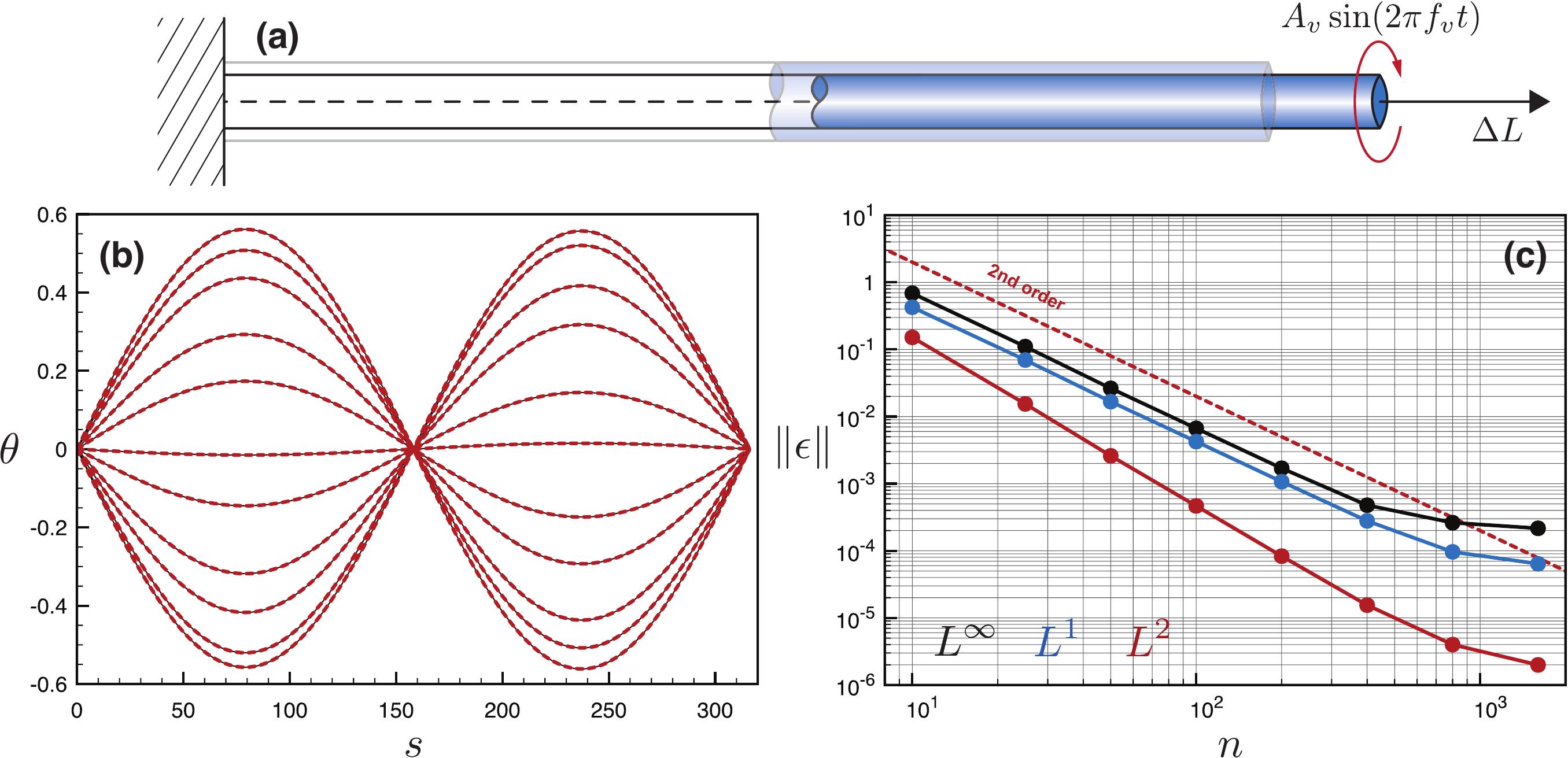}
\caption{\scriptsize{\textbf{Time-space convergence study for twist forced vibrations in a stretched rod.} (a) We consider a rod clamped at one end, forced to vibrate by applying the periodic couple $A_v\sin(2\pi f_v t)$ to the free end, and characterized by rest length $\hat{L}$ which is extended to a final length $L=e\hat{L}$. (b) Comparison between analytical $\theta$ (black lines) and numerical $\theta^n$ (red dashed lines) angular displacement with respect to the reference configuration along a stretched rod. Each red (numerical simulation) and black (analytical solution) line corresponds to the angular displacement along a rod discretized with $n=1600$ elements, and sampled at regular intervals $\Delta t=T_v/10$ within one loading period $T_v=f_v^{-1}$. (c) Norms $L^{\infty}(\epsilon)$ (black), $L^1(\epsilon)$ (blue) and $L^2(\epsilon)$ (red) of the error $\epsilon=\|\theta-\theta^n\|$ at different levels of time-space resolution are plotted against the number of discretization elements $n$. Here, the time discretization $\delta t$ is slaved by the spatial discretization $n$ according to $\delta t = 10^{-4} \delta l$ seconds. For all studies, we used the following settings: rod's density $\rho=10$~kg/m$^3$, Young's modulus $E=10^6$~Pa, shear modulus $G=2E/3$~Pa, shear/stretch matrix $\hat{\mathbf{S}}=\text{diag}(4G\hat{A}/3, 4G\hat{A}/3, E\hat{A})$~N, bend/twist matrix $\hat{\mathbf{B}}=\text{diag}(E\hat{I}_1, E\hat{I}_2, G\hat{I}_3)$~Nm$^2$, forcing amplitude $A_v=10^3$~Nm, forcing frequency $f_v=1$~s$^{-1}$, dilatation factor $e$=$1.05$, rest length $\hat{L}$=$\sqrt{E/\rho}/(ef_v)$~m, rest radius $\hat{r}$=$0.5$~m, simulation time $T_{\text{sim}}$=$2000$~s. We enabled dissipation in the early stages of the simulations, letting $\gamma$ decay exponentially in time to a zero value.}}
\label{fig:longwaveLoad_conv}
\end{center}
\end{figure}

Next we examine the problem of twisting waves generated in a rod axially stretched, as illustrated in Fig.~\ref{fig:longwaveLoad_conv}a. The coupling between stretching and the other dynamic modes tests the rescaling in terms of the dilatation factor $e$ of Eqs.~(\ref{eq:linmomentfinal}, \ref{eq:angmomentfinal}).
 
The rod is clamped at one end, stretched by the quantity $\Delta L = (e-1)\hat{L}$, and forced to angularly vibrate about the axial direction by applying at the free end the couple $A_v\sin(2\pi f_v t)$, where $A_v$ and $f_v$ are the corresponding forcing amplitude and frequency. In the case of no internal dissipation and constant circular sections, the induced standing angular wave $\theta(s,t)$ can be determined analytically. The dynamics of twisting yields the wave equation for the angular displacement \cite{Selvadurai:2000}
\begin{equation}
\frac{\partial^2 \theta}{\partial s^2}=\frac{1}{c_s}\frac{\partial^2 \theta}{\partial t^2},
\label{eq:twistWave}
\end{equation}
where $c_s=\sqrt{G/\rho}$ is the shear wave velocity, $G$ is the shear modulus, and $\rho$ is the density. By assuming a solution of the form $\theta(s,t)=\phi(s)\sin(2\pi f_v t)$, and by applying the boundary conditions $\phi(0)=0$ and $d_s\phi(0)=C_v/(\hat{I}_3 G)$ with $C_v$ twisting torque and $\hat{I}_3$ second area moment of inertia about the axial direction, we can solve Eq.~(\ref{eq:twistWave}) obtaining 
\begin{equation}
\theta(s,t) = \frac{A_v c_s}{2\pi f_v G \displaystyle\frac{\hat{I}_3}{e^2}}\cdot\frac{\sin\left(\displaystyle\frac{2\pi f_v e \hat{s}}{c_s}\right)}{\cos\left(\displaystyle\frac{2\pi f_v e\hat{L}}{c_s}\right)}\cdot\sin(2\pi f_v t).
\end{equation}

As can be seen in Fig.~\ref{fig:longwaveLoad_conv}b, our numerical method recovers the derived analytical solution for the twist angular displacement along the stretched rod. Moreover, the solver consistently exhibits a second order convergence in time and space given the error metric $\epsilon=\|\theta-\theta^n\|$, where $\theta^n$ represents our numerical solutions in the limit of refinement (Fig.~\ref{fig:longwaveLoad_conv}c). 

\section{Further friction validations}

\subsection{Validation of rolling friction model}
Here we validate the friction model introduced in Section \ref{sec:friction} on three test cases that can be analytically characterized. In all benchmarks we consider a rigid, unshearable, inextensible straight rod of mass $m$, length $L$, radius $r$, axial mass second moment of inertia $J$. The rod interacts with a surface characterized by static and kinetic friction coefficients $\mu_s$ and $\mu_k$, thus assuming isotropic friction.

\begin{figure}
\begin{center}
\includegraphics[width=\textwidth]{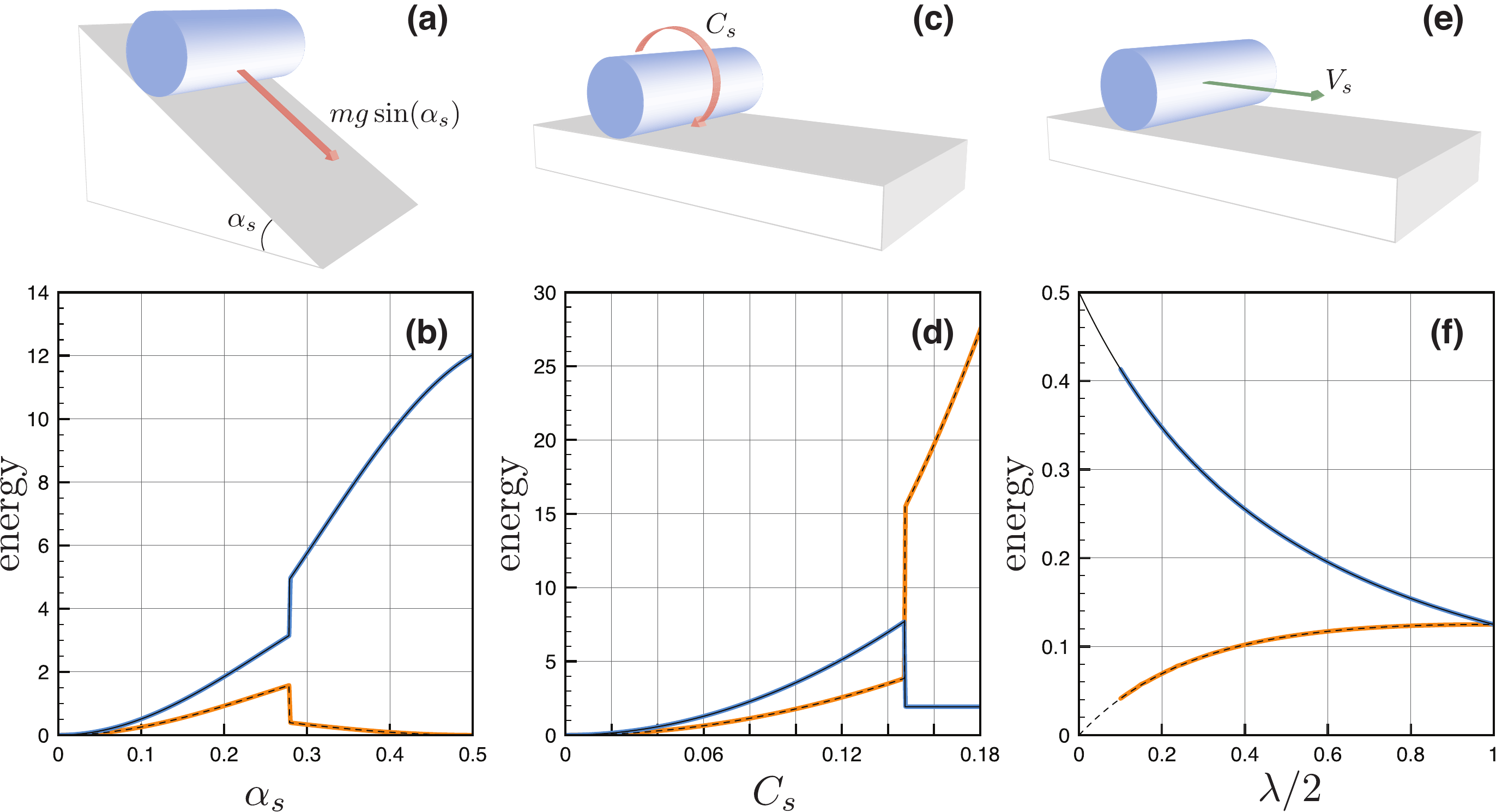}
\caption{\scriptsize{\textbf{Rolling static and dynamic friction.} (a) A rod initially at rest on a plane inclined of the angle $\alpha_s$, rolls or slips down with linear and angular velocities $v$ and $\omega$ due to its own weight $mg$. (b) Translational $E_T=mv^2/2$ (analytical solution - black line, numerical solution - blue line) and rotational $E_R=J\omega^2/2$ (analytical solution - dashed black line, numerical solution - orange line) energies are plotted against the angle $\alpha_s$. Settings: length $L=1$~m, radius $r=0.025$~m, mass $m=1$~kg, Young's modulus $E=10^9$~Pa, shear modulus $G=2E/3$~Pa, shear/stretch matrix $\mathbf{S}=10^4\cdot\pmb{1}$~N, bend/twist matrix $\mathbf{B}=\text{diag}(EI_1, EI_2, GI_3)$~Nm$^2$, dissipation constant $\gamma=10^{-6}$~kg/(ms), gravity $g=9.81$~m/s$^2$, static and kinetic friction coefficients $\mu_s=0.4$ and $\mu_k=0.2$, friction threshold velocity $v_{\epsilon}=10^{-4}$~m/s, ground stiffness and viscous dissipation $k_w=10$~kg/s$^2$ and $\gamma_w=10^{-4}$~kg/s, discretization elements $n=50$, timestep $\delta t = 10^{-6}$~s, simulation time $T=0.5$~s. (c)  Rod set in motion by the external couple $C_s$ on a horizontal plane. (d) Translational $E_T$ and rotational $E_R$ energies are plotted agains the couple $C_s$. Color scheme and settings identical to those of panel (b) except for the simulation time $T=1$~s. (e) Rod with initial velocity $V_s$ slows down due to kinetic friction until the no slip condition is reached. (f) Translational $E_T$ and rotational $E_R$ energies are plotted agains the relative mass moment of inertia ratio $\lambda/2$. Color scheme and settings identical to those of panel (b) except for the simulation time $T=2$~s, and the friction threshold velocity $v_{\epsilon}=10^{-6}$~m/s}.}
\label{fig:surfcefriction_app}
\end{center}
\end{figure}

In the first test the rod is initially at rest on a plane inclined of the angle $\alpha_s$, as depicted in Fig.~\ref{fig:surfcefriction_app}a. Due to the gravitational acceleration $g$ the rod starts rolling or slipping, depending on $\alpha_s$, down the plane. The linear $v$ and angular $\omega$ velocities of the filament, and therefore the corresponding translational $E_T=mv^2/2$ and rotational $E_R=J\omega^2/2$ energies can be analytically determined. 

By recalling Eq.~(\ref{eq:noslipforcerotation}), the force necessary to ensure rolling without slip takes the form $F_{\text{noslip}}=-F_{\parallel}/3=-mg\sin(\alpha_s)/3$. Given the maximum static friction force $F_s=\mu_s F_{\perp}=\mu_s m g \cos(\alpha_s)$ in the case $|F_{\text{noslip}}|\le|F_s|$ we have $a=(F_{\parallel}+F_{\text{noslip}})/m$. Therefore, at the time $T$ after releasing the rod, linear and angular velocities read, respectively, $v=aT$ and $\omega=\dot{\omega}T=(a/r)T$ expressing the no slip kinematic constraint between linear $a$ and angular $\dot{\omega}$ accelerations. On the contrary, if $|F_{\text{noslip}}|>|F_s|$ the rod starts slipping and linear and angular accelerations are no longer coupled, so that after the time $T$ we have $a=(F_{\parallel}-\mu_k F_{\perp})/m$, $v=aT$, $\omega=\dot{\omega}T=(\mu_k F_{\perp}r/J)T$. Therefore, translational and rotational energies as a function of $\alpha_s$ finally read
\[
E_T=
\begin{dcases}
	\frac{2mg^2T^2\sin^2(\alpha_s)}{9}& \text{if } |F_{\text{noslip}}|\le|F_s|\\
	\frac{mg^2T^2\left[\sin(\alpha_s)-\mu_k\cos(\alpha_s)\right]^2}{2}              & \text{if }  |F_{\text{noslip}}|>|F_s|
\end{dcases}
\]
and
\[
E_R=
\begin{dcases}
	\frac{2Jg^2T^2\sin^2(\alpha_s)}{9r^2}& \text{if } |F_{\text{noslip}}|\le|F_s|\\
	\frac{\mu_k^2m^2g^2r^2T^2\cos^2(\alpha_s)}{2J}              & \text{if }  |F_{\text{noslip}}|>|F_s|
\end{dcases}
\]
As can be noticed in Fig.~\ref{fig:surfcefriction_app}b, our numerical approach faithfully reproduces the derived analytical solution, accurately capturing the discontinuity at the transition between pure rolling and slipping.

The second test case of Fig.~\ref{fig:surfcefriction_app}c,d entails a rod set in motion by the external couple $C_s$ on a horizontal plane. Depending on the magnitude of the load the filament exhibits pure rolling or slipping motion. By recalling Eq.~(\ref{eq:noslipforcerotation}), the force necessary to ensure rolling without slip takes the form $F_{\text{noslip}}=2C_s/(3r)$. Given the maximum static friction force $F_s=\mu_s F_{\perp}=\mu_s m g$ in the case $|F_{\text{noslip}}|\le|F_s|$ we have $a=F_{\text{noslip}}/m$. Therefore, at the time $T$ after releasing the rod, linear and angular velocities read, respectively, $v=aT$ and $\omega=\dot{\omega}T=(a/r)T$ expressing the no slip kinematic constraint between linear $a$ and angular $\dot{\omega}$ accelerations. On the contrary, if $|F_{\text{noslip}}|>|F_s|$ the rod starts slipping and linear and angular accelerations are no longer coupled, so that after the time $T$ we have $a=\mu_k F_{\perp}/m$, $v=aT$, $\omega=\dot{\omega}T=J^{-1}(C_s-\mu_k F_{\perp}r)T$. Therefore, translational and rotational energies as a function of $C_s$ finally read
\[
E_T=
\begin{dcases}
	\frac{2T^2C_s^2}{9mr^2}			& \text{if } |F_{\text{noslip}}|\le|F_s|\\
	\frac{m\mu_k^2 g^2 T^2}{2}		& \text{if }  |F_{\text{noslip}}|>|F_s|
\end{dcases}
\]
and
\[
E_R=
\begin{dcases}
	\frac{2JT^2C_s^2}{9r^4m^2}		& \text{if } |F_{\text{noslip}}|\le|F_s|\\
	\frac{(C_s-\mu_k m g r)^2T^2}{2J}              & \text{if }  |F_{\text{noslip}}|>|F_s|
\end{dcases}
\]
As can be noticed in Fig.~\ref{fig:surfcefriction_app}d, again our numerical approach faithfully reproduces the derived analytical solution, accurately capturing the discontinuity at the transition between pure rolling and slipping.

In the last test case we consider a rod of initial velocity $V_s$ in the direction parallel to a horizontal plane and perpendicular to the filament axis (Fig.~\ref{fig:surfcefriction_app}e), and we vary the relative mass moment of inertia ratio $\lambda=2J/(mr^2)$. Initially the rod slips on the surface and gradually slows down, due to the effect of the kinetic friction, to a point for which linear $a$ and angular velocity $\omega$ meet the kinematic constraint $v_{eq}=r\omega_{eq}$ characteristic of pure rolling motion. The frictional force $F$ induces the torque $M=rF$, so that over a period of time $T$ we have $v=V_s-FT/m$ and $\omega=rFT/J$. By enforcing the no slip kinematic constraint $V_s-FT/m=r^2FT/J$, we have that $FT=V_s/(r^2/J+1/m)$. This allows us to directly compute $v_{eq}$ and $\omega_{eq}$ and the corresponding translational and rotational energies, yielding
\begin{equation}
E^{eq}_T=\frac{mV_s^2}{2}\frac{1}{(1+\lambda/2)^2},~~~~~~~E^{eq}_R=\frac{mV_s^2}{2}\frac{\lambda/2}{(1+\lambda/2)^2}.
\end{equation}
As depicted in Fig.~\ref{fig:surfcefriction_app}f, our numerical approach accurately reproduces the predicted energies as a function of $\lambda$.

\subsection{Validation of anisotropic longitudinal friction model}
\begin{figure}
\begin{center}
\includegraphics[width=\textwidth]{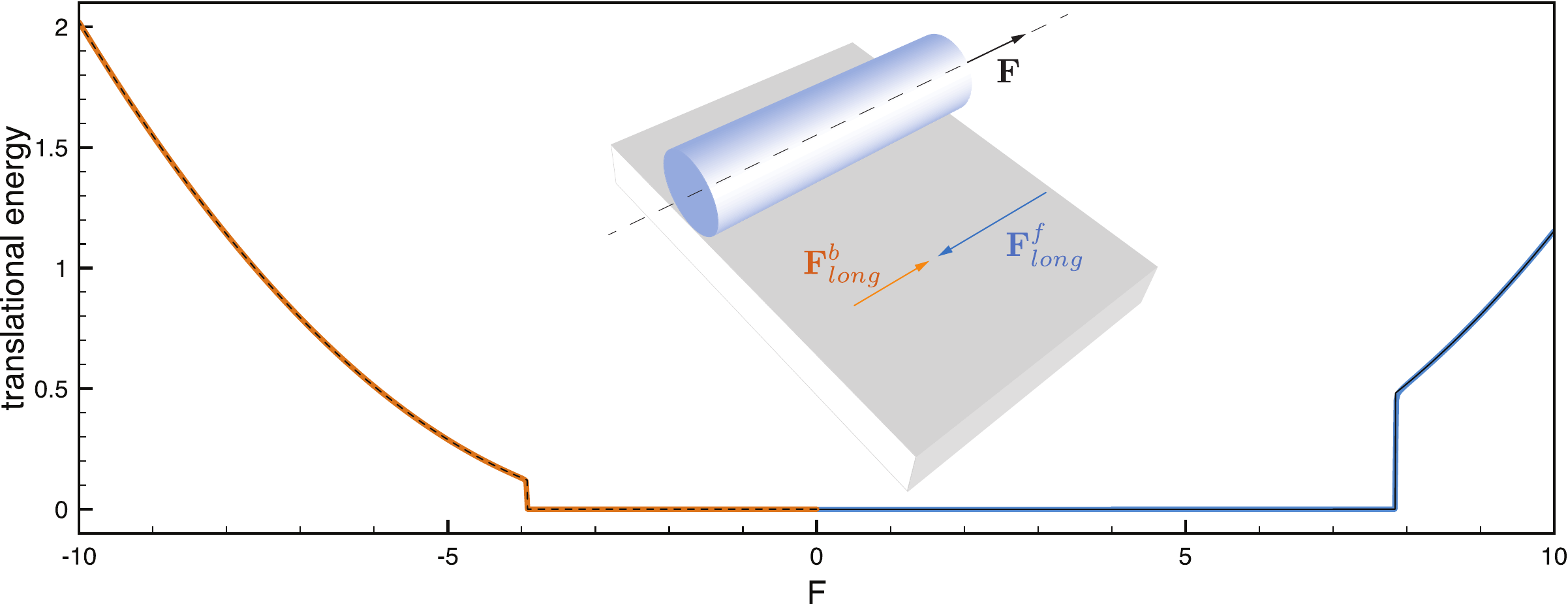}
\caption{\scriptsize{\textbf{Anisotropic static and kinetic longitudinal friction.} A rod initially at rest on a horizontal  plane is pulled longitudinally with force $\mathbf{F}$. The system is characterized by anisotropic friction so that resistance forces in the forward direction $\mathbf{F}^f_{long}$ are larger than in the backward direction $\mathbf{F}^b_{long}$. The plot illustrates the behavior of the total rod's translational energy $E_T$ as a function of the force $\mathbf{F}$ applied. Blue and solid black lines correspond to, respectively, simulated and analytical $E_T$ for a rod pulled forward. Orange and dashed black lines correspond to, respectively, simulated and analytical $E_T$ for a rod pulled backward. Settings: length $L=1$~m, radius $r=0.025$~m, mass $m=1$~kg, Young's modulus $E=10^5$~Pa, shear modulus $G=2E/3$~Pa, shear/stretch matrix $\mathbf{S}=10^4\cdot\pmb{1}$~N, bend/twist matrix $\mathbf{B}=\text{diag}(EI_1, EI_2, GI_3)$~Nm$^2$, dissipation constant $\gamma=10^{-6}$~kg/(ms), gravity $g=9.81$~m/s$^2$, static and kinetic forward friction coefficients $\mu^f_s=0.8$ and $\mu^f_k=0.4$, static and kinetic backward friction coefficients $\mu^b_s=0.4$ and $\mu^b_k=0.2$, friction threshold velocity $v_{\epsilon}=10^{-4}$~m/s, ground stiffness and viscous dissipation $k_w=10$~kg/s$^2$ and $\gamma_w=10^{-4}$~kg/s, discretization elements $n=50$, timestep $\delta t = 10^{-5}$~s, simulation time $T=0.25$~s.}}
\label{fig:surfcefrictionAxial_app}
\end{center}
\end{figure}

After validating our rolling friction model, we turn to its longitudinal counterpart. Here we consider a straight, rigid, inextensible and unshearable rod which is axially pulled or pushed with force $\mathbf{F}$ for a fixed period of time $T$ (Fig.~\ref{fig:surfcefrictionAxial_app}). Anisotrpy is captured through the forward $\mu_s^f$, $\mu_k^f$ and backward $\mu_s^b$, $\mu_k^b$ static and kinetic coefficients. For $\mathbf{F}\cdot \mathbf{t}\ge0$ the rod translational energy takes the form
\[
E^f_T=
\begin{dcases}
	0			& \text{if } |\mathbf{F}|\le|\mu^f_smg|\\
	\frac{T^2}{2m}(|\mathbf{F}|-\mu_k^fmg)^2		& \text{if }  |\mathbf{F}|>|\mu^f_smg|
\end{dcases},
\]
while for $\mathbf{F}\cdot \mathbf{t}<0$ we have
\[
E^b_T=
\begin{dcases}
	0			& \text{if } |\mathbf{F}|\le|\mu^b_smg|\\
	\frac{T^2}{2m}(|\mathbf{F}|-\mu_k^bmg)^2		& \text{if }  |\mathbf{F}|>|\mu^b_smg|
\end{dcases}.
\]
As can be seen in Fig.~\ref{fig:surfcefrictionAxial_app}, our numerical method reproduces the above theoretical predictions.

\subsection{Isotropic vs. anisotropic friction driven locomotion}
\label{sec:isovsani_app}
In this section we illustrate the effect of symmetry breaking via anisotropic friction in a system constituted by a soft filament interacting via surface friction with a solid substrate. If we consider isotropic friction and a specular muscular activation pattern by setting the control values $\beta_1 = \beta_4$ and $\beta_2=\beta_3$ and the wave number $2\pi/\lambda_m=0$ (see Section~\ref{sec:muscularActivity}), then the system is symmetric (up to inertial effects that can be neglected for small Froude numbers). Therefore, over any activation cycle the snake center of mass does not move. On the contrary, if we introduce anisotropy the snake will be able to slowly move (the capability to effectively move is impaired by the absence of the traveling gait component since $2\pi/\lambda_m=0$).

\begin{figure}
\begin{center}
\includegraphics[width=.95\textwidth]{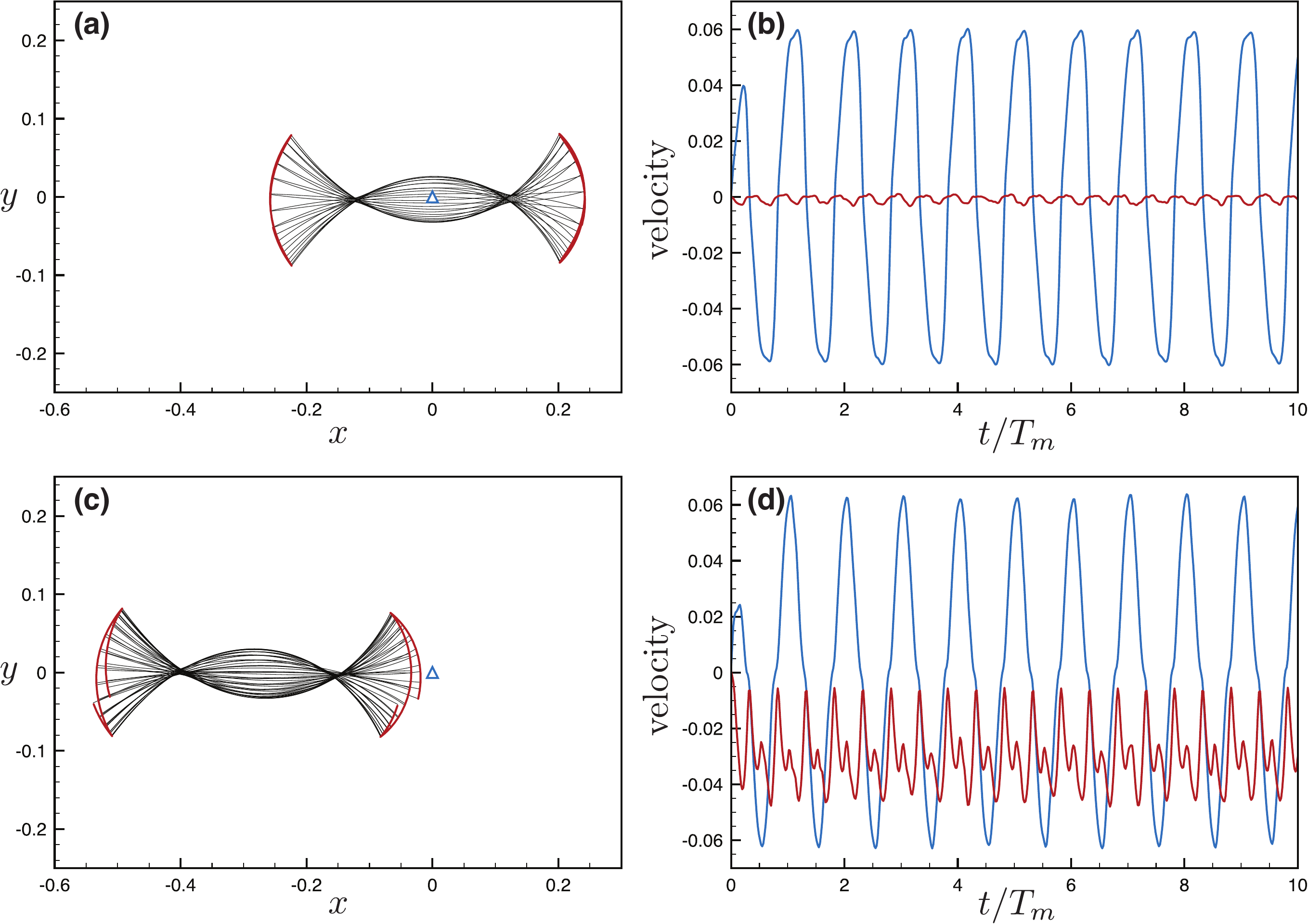}
\caption{\scriptsize{\textbf{Isotropic vs. anisotropic friction driven locomotion.} (a) Gait envelope computed over the 10th muscular activation cycle in the case of isotropic friction. The blue triangle denotes the location of the snake's center of mass at time $t=0$, reported as reference. (b) Lateral (blue) and forward (red) velocities as functions of time normalized by the activation period $T_m$ in the case of isotropic friction. (c) Gait envelope computed over the 10th muscular activation cycle in the case of anisotropic friction. The blue triangle denotes the location of the snake's center of mass at time $t=0$, reported as reference. (d) Lateral (blue) and forward (red) velocities as functions of time normalized by the activation period $T_m$ in the case of anisotropic friction. Settings: length $L=0.5$~m, radius $r=0.025$~m, mass $m=1$~kg, Young's modulus $E=10^7$~Pa, shear modulus $G=2E/3$~Pa, shear/stretch matrix $\mathbf{S}=10^5\cdot\pmb{1}$~N, bend/twist matrix $\mathbf{B}=\text{diag}(EI_1, EI_2, GI_3)$~Nm$^2$, dissipation constant $\gamma=10^{-1}$~kg/(ms), gravity $g=9.81$~m/s$^2$, static $\mu^f_s=0.2$, $\mu^r_s=\mu^f_s$, $\mu^b_s=\mu^f_s$ and kinetic $\mu^f_k=0.1$, $\mu^r_k=\mu^f_k$, $\mu^b_k=\mu^f_k$ friction coefficients in the isotropic case, static $\mu^f_s=0.2$, $\mu^r_s=2\mu^f_s$, $\mu^b_s=20\mu^f_s$ and kinetic $\mu^f_k=0.1$, $\mu^r_k=2\mu^f_k$, $\mu^b_k=20\mu^f_k$ friction coefficients in the anisotropic case, friction threshold velocity $v_{\epsilon}=10^{-8}$~m/s, ground stiffness and viscous dissipation $k_w=1$~kg/s$^2$ and $\gamma_w=10^{-6}$~kg/s, discretization elements $n=100$, timestep $\delta t = 10^{-5}$~s, muscular activation period $T_m=1$~s, wavelength $\lambda_m=\infty$, phase shift $\phi_m=0$, torque B-spline coefficients $\beta_{i=0,\dots,5}=\{0,10,15,15,10,0\}$~Nm.}}
\label{fig:IsoAnisotrpicFriction_app}
\end{center}
\end{figure}

As can be observed in Fig.~\ref{fig:IsoAnisotrpicFriction_app}, this prediction is captured by our numerical scheme, which accurately resolves the physical mechanisms at play. Moreover, this test shows once again how our methodology is robust in terms of numerical noise as no spurious displacements or rotations are generated in the symmetric case (Fig.~\ref{fig:IsoAnisotrpicFriction_app}a,b). 

\bibliographystyle{unsrtnat}

\end{document}